\documentclass[mnsc,sglanonrev]{ssrn}
\usepackage{fancyhdr}
\usepackage{eqndefns-left} 
\usepackage{mathtools}
\usepackage{extarrows}
\RequirePackage{tgtermes}
\RequirePackage{newtxtext}
\RequirePackage{newtxmath}
\RequirePackage{bm}
\RequirePackage{endnotes}

\OneAndAHalfSpacedXI

\usepackage{algorithm}
\usepackage{algpseudocode}
\usepackage{tikz}

\usepackage{enumitem}
\usepackage{caption}
\usepackage[labelfont=sf]{subcaption}
\usepackage{fix-cm}

\usepackage{natbib}
 \bibpunct[, ]{(}{)}{,}{a}{}{,}%
 \def\bibfont{\small}%
\usepackage{xcolor}
\usepackage[colorlinks=true,
            linkcolor=blue,
            citecolor=blue,
            urlcolor=blue]{hyperref}

\EquationsNumberedThrough    

\TheoremsNumberedThrough     
\ECRepeatTheorems  %

\renewcommand{\theARTICLETOP}{}

\makeatletter
\def\theARTICLEABSTRACT{%
\HOOKb
  \vspace*{18pt}
  \begin{center}
    \begin{minipage}[t]{0.85\textwidth} 
      \ABSfont\small 
      
       {\centering {\rmfamily\bfseries\fs.13.15. Abstract} \par}
      \vspace{8pt} 

      \noindent \theABSTRACT \endgraf
      
      \vskip12pt %
    \theKEYWORDS
    \theSUBJECTCLASS
    \theAREAOFREVIEW
    \theMSCCLASS
    \theORMSCLASS
  \noindent 
  \end{minipage}
  \end{center}
  \vspace*{0pt}
}
\makeatother

\begin{document}

\RUNAUTHOR{Qian, Mehra, and Liu}

\TITLE{The Economics of AI Supply Chain Regulation\thanks{
An earlier version of this paper, titled ``The Economics of Fine-Tuning for Large-Scale AI Models,'' was presented at WISE 2023, where it won the Best Student Paper Award.
}}

\ARTICLEAUTHORS{
\AUTHOR{Sihan Qian\thanks{School of Economics and Management, Tsinghua University. Email: \EMAIL{qsh22@mails.tsinghua.edu.cn}.} \quad
Amit Mehra\thanks{Naveen Jindal School of Management, The University of Texas at Dallas. Email: \EMAIL{amit.mehra@utdallas.edu}.} \quad 
Dengpan Liu\thanks{School of Economics and Management, Tsinghua University. Email: \EMAIL{liudp@sem.tsinghua.edu.cn}.}}
}
\date{December 17, 2025}
\ABSTRACT{%
The rise of foundation models has driven the emergence of AI supply chains, where upstream foundation model providers offer fine-tuning and inference services to downstream firms developing domain-specific applications. Downstream firms pay providers to use their computing infrastructure to fine-tune models with proprietary data, creating a co-creation dynamic that enhances model quality. Amid concerns that foundation model providers and downstream firms may capture excessive consumer surplus, along with increasing regulatory measures, this study employs a game-theoretic model involving a provider and two competing downstream firms to analyze how policy interventions affect consumer surplus in the AI supply chain. Our analysis shows that policies promoting price competition in downstream markets (i.e., \textit{pro–price–competitive policies}) boost consumer surplus only when compute or data preprocessing costs are high, while compute subsidies are effective only when these costs are low, suggesting these policies complement each other. In contrast, policies promoting quality competition in downstream markets (i.e., \textit{pro–quality–competitive policies}) always improve consumer surplus. We also find that under pro–price–competitive policies or compute subsidies, both the provider and downstream firms can achieve higher profits along with greater consumer surplus, creating a win–win–win outcome. However, pro–quality–competitive policies increase the provider's profits while reducing those of downstream firms. Finally, as compute costs decline, pro–price–competitive policies may lose their effectiveness, whereas compute subsidies may shift from ineffective to effective. These findings offer insights for policymakers seeking to foster AI supply chains that are economically efficient and socially beneficial.
}%
\maketitle
\pagestyle{fancy}
\fancyhf{}                
\fancyfoot[C]{\thepage}   
\renewcommand{\headrulewidth}{0pt}   
\renewcommand{\footrulewidth}{0pt}   

\newpage
\section{Introduction}
The rapid advancement of artificial intelligence (AI) is transforming economic sectors on an unprecedented scale \citep{sun2025effect}. The global AI market, valued at \$638.23 billion in 2024, is projected to grow at a compound annual growth rate (CAGR) of 19.2\% to reach \$3,680.47 billion by 2034 \citep{Precedence}. However, this profound growth introduces significant risks to consumer welfare, including deceptive claims about AI tools that lead consumers to make suboptimal purchasing decisions \citep{FTCPunishDeceptiveAIClaims}. 
This dual nature of AI, as both a powerful productivity tool and a potential source of risk, highlights the urgent need for effective regulatory oversight.
However, applying traditional regulatory frameworks to AI is problematic because the technology operates within a complex, multi-layered market structure, often conceptualized as an \textit{AI supply chain} \citep{AISupplyChain}. This structure is characterized by a distinct upstream-downstream dynamic: a small number of upstream providers develop large-scale AI models, also known as \textit{foundation models}, which are then adapted and deployed by a broad downstream market of firms for specific applications. The inherent interdependence of this structure renders existing regulatory frameworks inadequate, inevitably leading to substantial oversight gaps \citep{RegulationChallenging}. Therefore, safeguarding consumers and ensuring market efficiency necessitate a holistic regulatory approach tailored to address the unique interdependencies of the AI supply chain.

\subsection{Foundation Models}
To develop a holistic regulatory approach to the AI supply chain, it is essential to first gain a comprehensive understanding of its most fundamental component: the upstream segment, where developers create foundation models.
These models, built with billions of parameters and trained on massive datasets, exhibit remarkable capabilities in tasks such as logical reasoning and natural language processing \citep{DefineLargeScaleAIModel}.
Often hailed as a cornerstone of the next industrial revolution \citep{IndustryRevolution}, foundation models have driven major technology companies, including Microsoft, Amazon, and Google, to make substantial investments to secure dominance in the upstream market, with Google alone reportedly committing over \$100 billion \citep{GoogleInvestment}.\looseness=-1

Despite their impressive general-purpose performance, foundation models often fall short in specialized domains such as healthcare and law. 
These limitations stem primarily from the restricted access to domain-specific data, which is often highly confidential\citep{chen2026overview}.
Yet, organizations in these sectors, which have access to such data, generally lack the resources to develop foundation models independently due to the prohibitively high costs involved. These costs include not only massive computational resources, but also specialized technical expertise. 
For example, training a model at the scale of GPT-4 is estimated to cost over \$100 million, making in-house development infeasible for most organizations \citep{GPT-4}.

\subsection{Fine-tuning}
\begin{figure}
\FIGURE
{\includegraphics[width=0.5\textwidth]{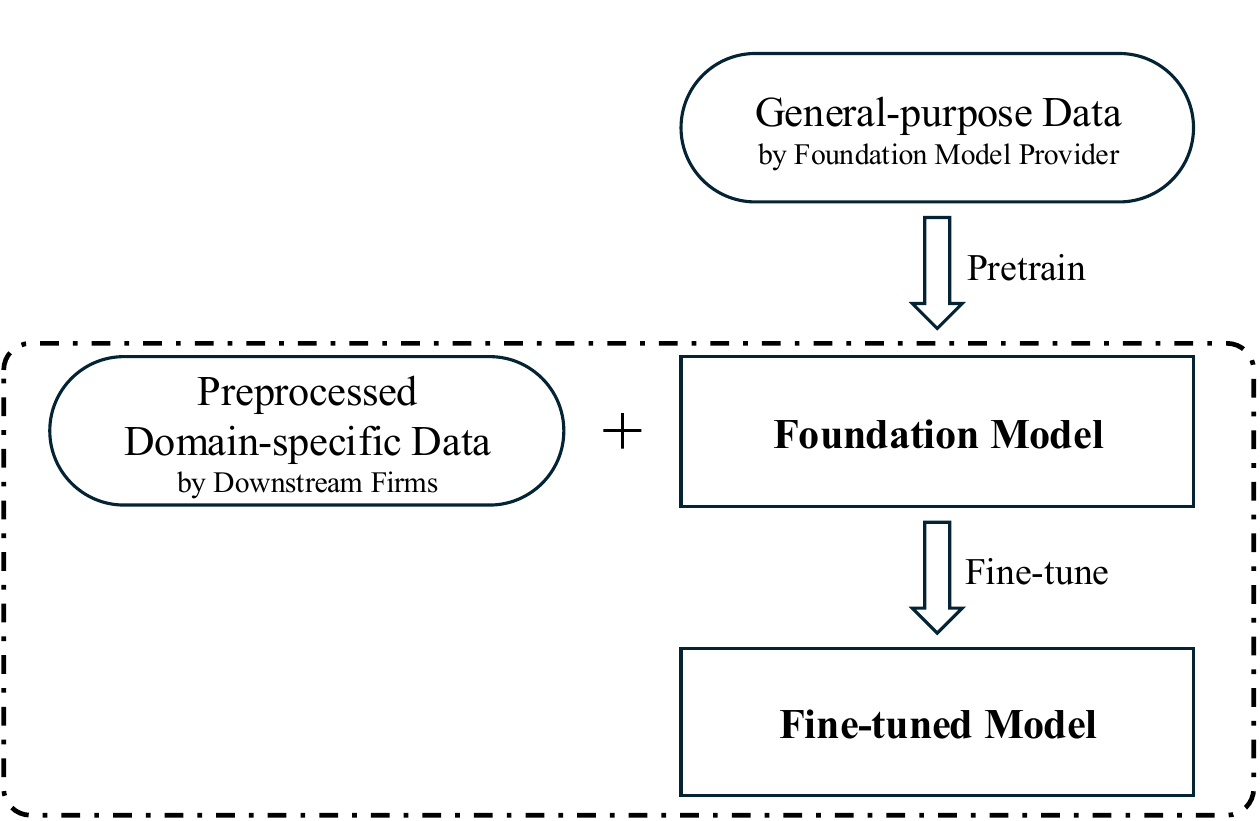}}
{Creation of a fine-tuned model
\label{fig:fine-tuning}}
{}
\end{figure}

To tackle these challenges, a co-creation process has emerged within a comprehensive AI ecosystem, bringing together upstream foundation model providers and downstream firms. As shown in Figure~\ref{fig:fine-tuning}, adapting a model for downstream applications typically requires two key training stages. The first stage is pretraining, which involves building a foundation model using general-purpose data \citep{PretrainIsOnGeneralPurposeData}. Pretraining is primarily conducted by leading technology companies such as Google and OpenAI \citep{Pretrain}. \looseness=-1

The second stage, known as fine-tuning, is typically performed by downstream firms and consists of two critical steps. The first step is preprocessing domain-specific data to remove noise, errors, and inconsistencies\footnote{For example, the fine-tuning for CoCounsel was based on a curated dataset of approximately 30,000 legal questions, refined over six months by a multidisciplinary team of lawyers, domain experts, and AI engineers \citep{DataPreprocessingExample}.} \citep{DeDuplicationInFineTuningData}. Once the data is cleaned, it is used to retrain the foundation model, yielding a fine-tuned version tailored to specific downstream applications. Unlike the resource-intensive pretraining phase, fine-tuning requires significantly less computational power, making it a feasible option for most firms \citep{govuk2024aifoundation}. This reduced computational cost enables smaller companies to take advantage of the models developed by large technology companies and adapt them to their specialized needs \citep{AI}. Fine-tuning is widely regarded as essential for embedding complex domain-specific behavioral patterns in AI models \citep{OpenAI}. Supporting this, a 2024 report by PricewaterhouseCoopers (PwC) finds that organizations using fine-tuned models experienced a 37\% increase in the accuracy of AI-generated content \citep{90PercentUsegFine-tuning}. The same report predicts that by 2030, 90\% of companies will have deployed at least one fine-tuned large language model. 

To protect the proprietary nature of the foundation model's parameters, model providers do not share the models directly with downstream firms. Instead, firms are required to upload clean, domain-specific data, which is obtained by preprocessing their datasets, a process that incurs what are known as \textit{data preprocessing costs}, to the provider's infrastructure for fine-tuning. Given that fine-tuning requires significant computational resources, providers also face \textit{compute costs}. In exchange, they charge a \textit{fine-tuning fee}, typically based on the volume of preprocessed data firms use on the provider's infrastructure. Once the model has been fine-tuned, it is hosted on the provider's infrastructure, and firms are charged an \textit{inference fee}, usually based on the volume of usage data (both input and output), to access the model for downstream applications \citep{OpenAIPricing}. In this way, foundation model providers generate two primary revenue streams: one from fine-tuning fees and the other from inference fees.

\subsection{The AI Supply Chain}
\label{AI Supply Chain}
\begin{figure}
\FIGURE
{\includegraphics[width=0.4\textwidth]{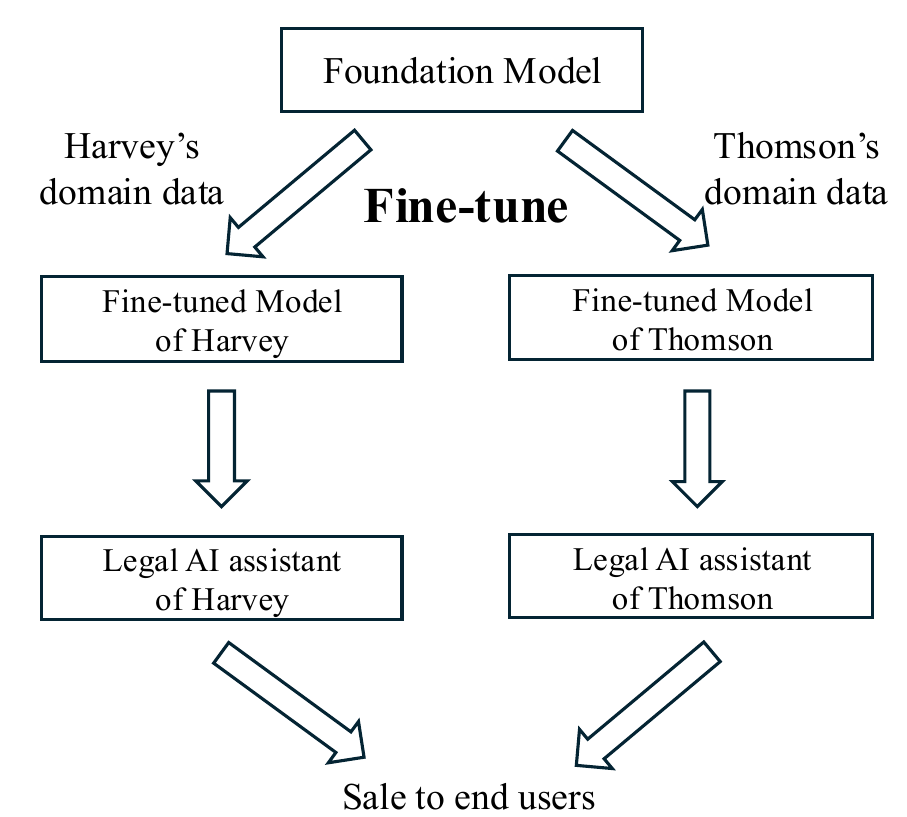}}
{Example of an AI supply Chain
\label{fig: AI Supply Chain}}
{}
\end{figure}

The central role of fine-tuning in AI development has led to the emergence of the AI Supply Chain \citep{AISupplyChain}, where foundation model providers offer general-purpose models that downstream firms fine-tune for specialized applications and subsequently market to end users. 
A real-world example of this AI supply chain is illustrated in Figure~\ref{fig: AI Supply Chain}. 
For instance, law firms such as Harvey and Thomson Reuters are the leading players in developing AI-powered legal assistants \citep{LeadingAILawFirms}. 
Both firms fine-tune OpenAI foundation models using their proprietary legal data \citep{Harvey, ThomsonReuters}. Using these fine-tuned models, they have created and commercialized advanced AI legal assistants that help attorneys with tasks such as drafting legal documents and managing complex litigation \citep{Harvey}.


This novel form of supply chain has several distinctive characteristics. First, the foundation models are fine-tuned through a co-creation process involving both the foundation model provider and the downstream firm. This approach to product development contrasts sharply with traditional supply chains, where manufacturers typically handle all aspects of product design and development. Even in collaborative supply chains, such as those that involve vendor–client partnerships for enterprise systems such as ERP \citep{demirezen2020two, gupta2023worse}, the co-creation process typically occurs independently between vendors and clients. In contrast, the fine-tuning of foundation models entails a mutually dependent form of co-creation: upstream providers supply not only the base foundation models but also the computational infrastructure necessary for fine-tuning, while downstream firms contribute domain-specific data and expertise to adapt these models for specialized applications. Second, the provision of computational resources for fine-tuning creates a revenue stream for foundation model providers during the co-creation process, which represents a unique form of revenue generation compared to traditional settings, where manufacturers generate revenue only through wholesale prices. Third, the fine-tuned models are customized versions of foundation models deployed through a Software-as-a-Service (SaaS) framework. Traditionally, software customization occurs in on-premise settings, where clients adapt the software locally; in such cases, SaaS-style pricing models are typically not applied. Hence, the integration of extensive customization with a SaaS deployment model constitutes a distinctive feature of our context.

\subsection{Motivation and Research Questions}
The rapid diffusion of AI-enabled products has introduced substantial risks to consumer welfare, particularly by shifting power from consumers to firms. 
As Nobel laureate Daron Acemoglu cautions, the use of AI technologies and data increases companies' power over consumers, with important distributional implications that enable AI-intensive companies to capture a larger share of consumer surplus~\citep{acemoglu2021harms}. \looseness=-1

These concerns are compounded by the fact that AI-enabled products introduce fundamentally new types of services for consumers (e.g., AI-based legal assistants), yet assessing their quality remains inherently difficult \citep{AIProductQualityHardToAssess2}.
For example, consumers cannot easily anticipate the extent of hallucinations, which can significantly undermine the reliability of the output.
Legal experts caution that AI systems remain susceptible to errors, biases, and malfunctions, requiring professionals to be aware of their limitations \citep{QualityTransparencyProblem}.
More concerningly, some companies have exploited the hype around AI, exaggerating the capabilities of their products and further complicating consumers’ ability to assess quality.
For example, DoNotPay marketed itself as the ``world’s first robot lawyer,'' falsely claiming that it could replace human legal expertise, only to fall short of those promises \citep{FTCPunishDeceptiveAIClaims}.
This lack of quality transparency enables downstream firms to soften competition, potentially leading to subpar offerings and a decline in consumer welfare.
In a similar vein, challenges also arise in pricing: Opaque and highly differentiated pricing schemes hinder straightforward comparisons. As \citet{Dreyer} notes, reduced comparability undermines competitive pressure and tends to drive prices higher.


Given these issues, a natural regulatory response to AI supply chains is to implement pro-competitive policies that strengthen price or quality competition in downstream markets \citep{austria2024ai}. Such approaches have already been successfully applied in the broader digital economy. For instance, to promote price competition, the U.S. Federal Trade Commission (FTC) required Ticketmaster to display the full ticket price upfront during the purchase process, thereby improving price transparency \citep{Ticketmaster}. Similarly, in a quality competition context, the FTC alleged that Fashion Nova, a U.S.-based online fast-fashion retailer, misled consumers by suppressing negative reviews, underscoring the importance of truthful product quality disclosure \citep{FashionNova}. Crucially, some of these established regulatory tools have already been adapted to AI-enabled product markets. For example, the FTC has begun to require companies to refrain from making misleading claims about the capabilities of consumer-facing AI products \citep{FTCPunishDeceptiveAIClaims2,FTCAIGuideline}. Ultimately, these measures contribute to a broader competition policy framework by limiting misleading practices that distort consumer choice and weaken competitive pressures \citep{DeceptiveActReduceCompetition}. \looseness=-1

Another regulatory strategy targets upstream foundation model providers in the AI supply chain, aiming to enhance consumer surplus by subsidizing the fine-tuning process and incentivizing the development of higher-quality products and services in downstream markets. A notable example is the annual commitment by the Beijing Municipal Government of 100 million RMB in computing power subsidies to firms engaged in large-scale AI model services \citep{BeijingSubsidy}. 
Similar programs have been introduced by municipal governments in Shenzhen, Shanghai, Chengdu, and several other Chinese cities. By offsetting a significant portion of the compute costs for fine-tuning, these subsidies help lower barriers to producing higher-quality AI-enabled offerings. \looseness=-1

However, the implications of such regulatory measures in the AI supply chain are not straightforward and may differ from those seen in traditional supply chain models, due to the unique co-creation dynamics inherent in this intricate ecosystem. In conventional supply chains, pro-competitive policies that lower retail prices typically increase consumer surplus without diminishing the manufacturer’s incentive to maintain product quality. In contrast, in the AI supply chain context, a reduction in downstream prices can erode downstream firms' marginal returns, diminishing their incentives to invest in fine-tuning and ultimately lowering overall product quality. This interdependence between pricing, incentives, and quality gives rise to a welfare trade-off absent in conventional market structures.

This trade-off highlights the need to evaluate not only whether regulatory interventions enhance consumer welfare but also the mechanisms through which they operate and the conditions under which they may inadvertently discourage quality investment. Understanding these mechanisms is essential because the welfare implications of a policy depend on how it shapes firms’ incentives: Measures that appear pro-competitive can, under certain conditions, diminish firms' incentives to invest in quality, ultimately undermining overall welfare. 
Recent debates on AI regulation reveal growing gaps in adapting existing regulatory frameworks to complex AI supply chains, underscoring the importance of understanding how different policy instruments perform in such environments \citep{RegulationChallenging}.
These considerations motivate our central research questions: \textit{How do regulatory approaches that promote competition in downstream markets or provide compute subsidies impact consumer surplus within the AI supply chain, and which policy instrument proves most effective under varying conditions?} Addressing these questions is crucial to ensure that regulatory efforts ultimately promote, rather than compromise, welfare in AI-enabled markets.

The success of any regulatory policy ultimately depends on its acceptance by industry participants. Therefore, it is crucial to understand how pro-competitive policies and compute subsidies impact the profitability of both foundation model providers and downstream firms. Policies that increase consumer surplus while simultaneously improving the profitability of providers and downstream firms are more likely to be adopted, as they create a ``win–win–win'' outcome in which all stakeholders benefit. Conventional wisdom suggests that increased competition erodes downstream firms' market power, thereby reducing their profitability \citep{matsushima2006industry}. However, \citet{tyagi1999effects} provides an alternative view, showing that increased downstream rivalry can induce price adjustments by upstream providers, thereby easing competitive pressures among downstream firms and mitigating the need for aggressive pricing. This suggests that the relationship between downstream competition and firm profitability is theoretically ambiguous. In AI supply chains, where downstream firms co-create value through fine-tuning, the dynamics are likely even more complex. Furthermore, strategic interactions between upstream providers and downstream firms make the effects of compute subsidies on their profitability uncertain, as changes in incentives from one party can influence the other party's strategic responses. To address these uncertainties, we pose our second research question: \textit{How do pro-competitive policies and compute subsidies influence the profitability of both foundation model providers and downstream firms within the AI supply chain?}

While evaluating the immediate effects of regulatory policies is important, it is equally crucial to recognize how these policies must evolve over time to respond to the shifting dynamics of AI markets.
This also represents a key practical challenge for policymakers due to the rapid and continuous evolution of AI technologies \citep{AIFastEvolving}.
A key driver of these changes is the steady decline in compute costs, largely driven by advancements in graphics processing units (GPUs), which play a crucial role in fine-tuning AI models \citep{GPUIsComputingPower,FineTuningNeedGPU}. Lower compute costs enable providers to reduce fine-tuning prices, benefiting downstream firms. However, prior research suggests that falling IT costs can intensify competition over product quality, potentially eroding downstream profitability \citep{demirhan2005information}. Consequently, the overall impact of declining compute costs on downstream profitability remains uncertain. In AI supply chains, where downstream firms co-create value through fine-tuning, these dynamics may further influence incentives to invest in quality and, ultimately, consumer surplus. This raises important policy considerations: Regulatory interventions that are effective under current compute cost structures may need to be adjusted as costs decline in order to maintain their intended effects on consumer surplus and firm incentives. To examine the implications of declining compute costs, we pose our third research question: \textit{How do declining compute costs impact consumer surplus and the profitability of both foundation model providers and downstream firms within the AI supply chain, and how should regulatory policies evolve to ensure welfare-enhancing outcomes?}

\subsection{Key Findings}
To address these research questions, we develop a game-theoretic model in which a foundation model provider offers both fine-tuning and inference services to two competing downstream firms. These firms, in turn, leverage these services to develop and sell AI-powered products to consumers. Our analysis yields several key insights. 
First, we show that neither policies promoting price competition in downstream AI markets nor compute subsidies always effectively enhance consumer welfare.
Nevertheless, these two policy strategies are complementary, and their effectiveness depends on the underlying cost conditions, which relate to both the foundation model provider's compute costs and downstream firms' data preprocessing costs, reflecting the co-creation nature of the AI supply chain.
Specifically, policies that promote price competition in downstream AI markets are effective only when compute or data preprocessing costs are relatively high, whereas compute subsidies are effective only when these costs are relatively low. 
In contrast, policies that promote quality competition in downstream AI markets always enhance consumer welfare, making it a robust and broadly effective regulatory approach.

Second, contrary to conventional wisdom, we find that policies that promote price competition among downstream firms can increase their profits while reducing the foundation model provider's profits.
Moreover, both policies promoting price competition and compute subsidies can, in some cases, generate a ``win–win–win'' outcome in which all stakeholders benefit.
In contrast, policies promoting quality competition consistently reduce downstream firms' profits while increasing the provider's profits, offering no such ``win--win--win'' potential.

Third, as GPU technology advances and compute costs decline over time, both the consumer surplus and provider profits increase, but downstream firms' profits may decrease.
Furthermore, declining compute costs alter the effectiveness of policy interventions.
Policies promoting price competition may shift from effective to ineffective, while compute subsidies may transition from ineffective to effective. In contrast, policies that promote quality competition remain consistently effective.

The remainder of this paper is organized as follows. Section 2 reviews the relevant literature and situates our work within existing research. Section 3 presents the modeling framework underlying our analysis, while Section 4 develops formal models to examine the optimal decision-making processes of both the foundation model provider and the downstream firms. Section 5 evaluates the effectiveness of various regulatory interventions in AI markets, and Section 6 examines the impact of pro-competitive policies and compute subsidies on the profitability of key stakeholders. Section 7 explores how declining compute costs influence stakeholder profitability, consumer surplus, and the effectiveness of these regulatory policies. Finally, Section 8 concludes with a discussion of theoretical contributions and managerial implications.

\section{Literature Review}
Our study relates primarily to three streams of literature: (i) AI Regulation, (ii) Software as a Service (SaaS), and (iii) Implications of Declining Information Technology (IT) Costs. 
Below, we briefly review the relevant literature and highlight our contributions to each stream, with Figure 3 summarizing these works and illustrating the positioning of our study.

\subsection{AI Regulation}
Our paper contributes to the recently emerging literature on the regulation of AI. 
Prior work has analyzed the risks posed by AI, highlighting the urgent need for regulatory oversight. For example, \cite{acemoglu2021harms} highlights the broad social and economic harms associated with AI, while \cite{cajueiro2024comprehensive} offer a comprehensive review of the challenges involved in the governance of rapidly evolving AI technologies. Empirically, \cite{assad2024algorithmic} demonstrate that algorithmic pricing can facilitate collusion and weaken competition, underscoring regulators’ concerns about the widespread adoption of AI-powered pricing tools.
Another stream of literature proposes or examines specific regulatory approaches. \cite{mahari2024regulation} introduce a Regulation-by-Design paradigm, which integrates regulatory objectives directly into AI system architectures. Other studies investigate the unintended effects of particular interventions \citep{vijairaghavan2025consequences}. For example, \cite{mohammadi2024regulating} find that policies designed to improve AI explainability can unintentionally worsen outcomes for both consumers and firms, while \cite{wang2023algorithmic} show that policies requiring full transparency of AI algorithms can inadvertently harm end users.

Unlike these studies, we examine two regulatory approaches that have not yet been analyzed in the AI context, namely pro-competitive policies and compute subsidies. 
More importantly, we incorporate the multi-layer structure of the AI market, specifically the AI supply chain, a factor that can significantly influence policy effectiveness \citep{RegulationChallenging} but has been largely neglected in the literature.
By modeling the strategic interactions between upstream foundation model providers and downstream firms, our study offers a systematic framework for evaluating the implications of regulatory policies in the AI supply chain.



\subsection{SaaS}
Because fine-tuning and inference services can be viewed as forms of SaaS for AI \citep{WhySaaS}, our study is closely related to the literature on SaaS  \citep{yang2021cloud, sun2023enemy, jiang2019optimal}.
Competition among IT resource vendors has been extensively studied in this context. For example, \citet{Kim2009economic} examine quality competition between SaaS vendors and show that risk-sharing contracts can incentivize vendors to improve product quality. \citet{ma2014competition} investigate both price and quality competition, finding that high client switching costs can ultimately drive a vendor out of the market. Similarly, \citet{guo2018model} and \citet{zhang2020cloud} analyze the competitive dynamics between perpetual software vendors and SaaS providers, highlighting how factors such as quality improvement rates and network effects shape SaaS competitiveness. While this body of work provides valuable insights into vendor-side competition, it largely overlooks the competitive dynamics among downstream firms that use SaaS products. In particular, little attention has been paid to how competition among downstream users shapes SaaS vendors' pricing strategies. Our study addresses this gap by examining the interplay between vendor pricing (specifically, by foundation model providers) and downstream market competition. In doing so, we contribute novel insights into how the structure of downstream markets can influence upstream pricing decisions, with implications for both market participants and policymakers.

Prior research has primarily focused on customizable software services \citep{ma2015analyzing,zhang2020cloud}, with a particular emphasis on modifiable off-the-shelf (MOTS) products. These solutions are typically installed on-premises and offered at fixed prices, rather than through usage-based or on-demand billing, which is a distinctive characteristic of SaaS. In contrast, traditional SaaS models offer limited customization because multiple customers share a single software instance, thereby restricting access to source code modifications \citep{sun2008software,muller2009customizing, ma2014competition, guo2009survey, ma2015analyzing, zhang2020cloud}. While SaaS platforms often include configuration options, these are generally insufficient to accommodate advanced or domain-specific customization needs \citep{Schneier2015,zhang2020cloud,Karanika2013}. The emergence of large-scale AI models introduces a fundamental shift in the nature of service customization. Through fine-tuning, firms can retrain foundation models on proprietary data, enabling parameter adjustments without requiring access to or modification of the underlying source code. This capability fosters a new form of co-creation, wherein service quality reflects both the capabilities of the vendor's foundation model and the unique data contributed by downstream firms. This interdependence gives rise to strategic dynamics that diverge significantly from those observed in traditional SaaS settings. Despite these shifts, the strategic implications of customizable SaaS remain underexplored in the existing literature. Our study addresses this gap by examining the evolving interactions between SaaS vendors (i.e., foundation model providers in our context) and downstream firms within this emerging paradigm.

\begin{figure}
\FIGURE
{\includegraphics[width=0.8\textwidth]{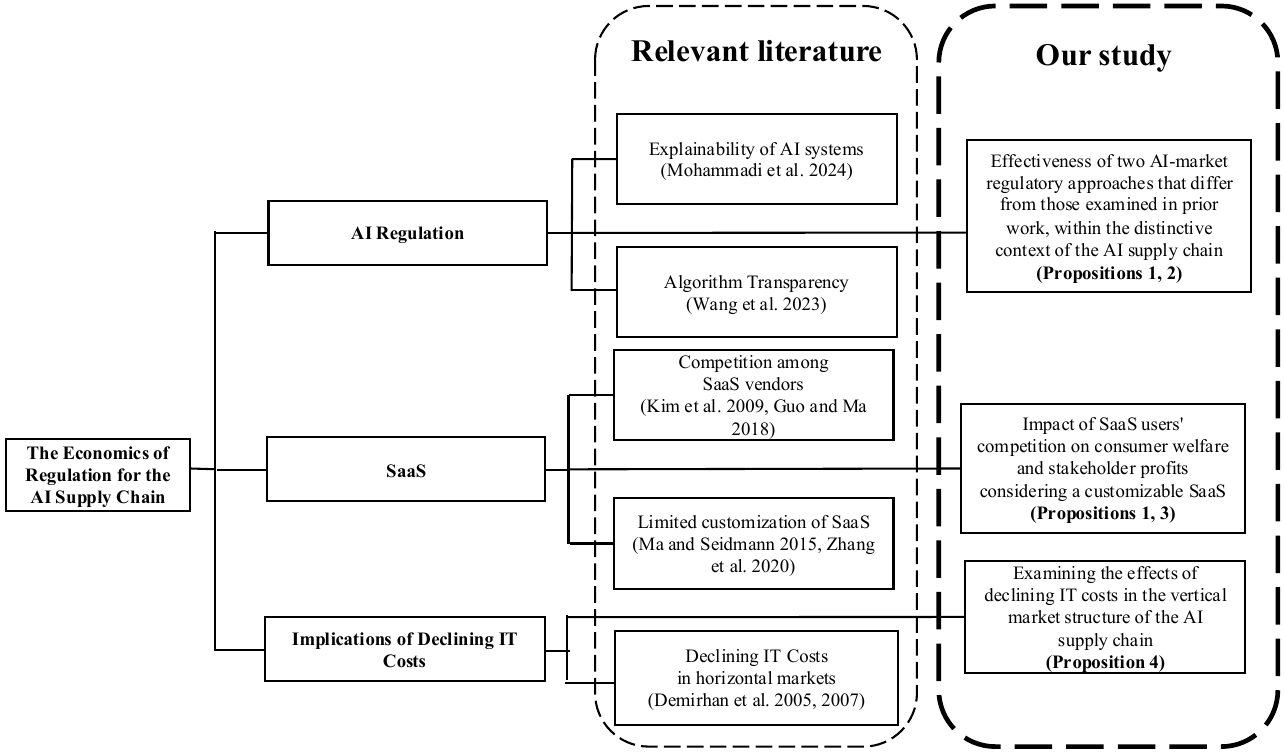}}
{Relevant Literature and the Positioning of Our Study
\label{fig:research position}}
{}
\end{figure}

\subsection{Implications of Declining IT Costs}
Given the continued decline in the costs of AI computing, our paper also contributes to the broader literature on declining IT costs. Prior research in this area has primarily focused on the implications of declining IT costs for firm profitability in competitive, consumer-oriented markets. For example, \cite{demirhan2005information,demirhan2007strategic} examine strategic IT investments and show that whether firms can benefit from declining IT costs depends on factors such as price sensitivity and user switching costs. Similarly, \cite{bohlmann2002deconstructing} analyze the sequential entry of firms into downstream markets and empirically show that the negative effect of declining IT costs on the profit of the pioneer is more pronounced in product categories where quality is a key differentiator.

In contrast to prior work, our study examines the effects of declining IT costs within a vertical market structure, specifically the AI supply chain that includes upstream providers, downstream firms, and end consumers. While earlier research has primarily focused on horizontal market settings, where interactions occur only between downstream firms and consumers, our model captures the strategic implications of vertical interactions throughout the supply chain. Moreover, while prior studies typically attribute IT costs to downstream firms, our analysis highlights the strategic impact of declining compute costs incurred by upstream providers.

\section{Modeling Framework}
We analyze a market scenario involving a foundation model provider (hereafter referred to as the \textit{provider}) and two competing downstream firms (denoted by \textit{firm $i$}, where $i\in\{1,2\}$). To ground the analysis in a real-world example, consider law firms such as Harvey and Thomson Reuters, which are prominent competitors in the provision of AI-powered legal assistants \citep{LeadingAILawFirms}. Both firms have fine-tuned OpenAI’s GPT model to develop legal assistants. This fine-tuning process enables legal assistants to deliver highly relevant and accurate results, thereby enhancing service quality and contributing significantly to firm profitability \citep{Harvey}. Such scenarios, where competing firms rely on the same foundation model provider, are increasingly common in practice \citep{LegalTechLLM}.

Next, we define the profit functions for each stakeholder and outline the game's sequence. 

\subsection{Foundation model provider}
To help firms leverage the foundation model, the provider offers two key services: fine-tuning and inference. Both services are essential for firms to deliver AI-powered products, such as AI-driven legal assistants, to end users, including individuals and enterprises. 

In the fine-tuning process, firms supply domain-specific data to the provider, who utilizes its computing infrastructure to adapt the foundation model, thereby enhancing its domain-specific performance. According to \citet{OpenAI}, the performance of a fine-tuned model improves with the volume of data provided. To quantify this relationship, we define a \textit{data unit} as the amount of data required to enhance the quality of model by one unit (e.g., 1,000 tokens\footnote{Tokens refer to words, character sets, or combinations of words and punctuation input or generated by large language models (LLMs).}). Additionally, we define \textit{data volume} as the volume of data used for fine-tuning, denoted by $V_i$. Hence, if firm $i$ provides $V_i$ data units for fine-tuning, the corresponding improvement in model quality is $V_i$. Therefore, the quality of the fine-tuned model can be represented as $q_i=\alpha+V_i$, where $\alpha$ denotes the base quality of the foundation model. Since fine-tuning is performed on the provider’s computing infrastructure, it incurs rising computational costs as the volume of training data increases. These costs are modeled using the quadratic expression $c_F V_i^2$, where $c_F$ is the cost coefficient associated with fine-tuning. The convex cost structure arises because the fine-tuning of large-scale AI models depends on distributed computing architectures \citep{GPUCluster1, GPUCluster2}. As the volume of data increases, additional GPU nodes are required, thereby increasing communication overhead between GPUs. This, in turn, creates computational bottlenecks that reduce GPU efficiency and cause costs to grow non-linearly \citep{narayanan2021efficient, zhao2021bridging}. To provide fine-tuning services, the provider sets a per-unit price of $p_F$ for the data used in this process. For example, OpenAI charges \$6 per million tokens to fine-tune its Davinci model, which is part of the GPT series \citep{OpenAIPricing}. Consequently, firm $i$ pays a total of $p_F V_i$ to the provider.

The fine-tuned model is deployed and hosted on the provider's computing infrastructure. To access the model, firms utilize the provider's inference service. During inference, firms submit queries to the fine-tuned model, which processes the input, generates responses, and returns the results. The provider typically charges firms on a pay-per-use basis for inference, with a price of $p_I$ per unit of data usage, which includes both the input data and the output generated by the model. For example, OpenAI charges \$12 per million tokens for inference on its Davinci model \citep{OpenAIPricing}. Since downstream firms develop AI-powered products based on the fine-tuned model, each consumer interaction with the product triggers a request for the inference service. For example, each query submitted by a lawyer to an AI-powered legal assistant invokes the fine-tuned model to generate a response, thereby incurring inference usage. For simplicity, we assume that answering a consumer's query requires one unit of data usage for inference. Therefore, the total data usage for inference is equal to the consumer demand, denoted by $d_i$. Consequently, firm $i$ pays $p_I d_i$ to the provider for inference services.

The objective of the foundation model provider is to maximize total profits by strategically setting prices for both fine-tuning and inference services. The provider’s profit function is given by:
\begin{equation}
\label{foundation model provider profits}
\pi_M= \sum_{i=1}^{2} \left( p_I d_i + p_F V_i - c_F V_i^2 \right).
\end{equation}
In this expression, the first term ($p_I d_i$) represents revenue generated from the inference service, where $d_i$ denotes the inference demand for firm $i$. The second term ($p_F V_i$) corresponds to revenue from fine-tuning services, while the third term ($c_F V_i^2$) captures the computational costs associated with fine-tuning.

\subsection{Firms}
Firms must preprocess their domain-specific data to address potential problems in raw data and ensure its suitability for fine-tuning \citep{DeDuplicationInFineTuningData}. For example, de-duplication, one of the key preprocessing steps, involves identifying and removing duplicate data points \citep{DeDuplicationInFineTuningData}. 
Although essential, data preprocessing represents a significant burden for firms. For instance, Thomson Reuters reportedly invested 4,000 hours of effort from a multidisciplinary team of lawyers, domain experts, and AI engineers to curate and refine 30,000 legal questions for fine-tuning \citep{DataPreprocessingExample}. The workload associated with preprocessing increases with the volume of data used for fine-tuning ($V_i$), as larger preprocessed datasets require more raw data and greater processing complexity. For example, the de-duplication process involves pairwise comparisons and thus exhibits quadratic computational complexity with respect to the dataset size \citep{munappy2022data}. Accordingly, we model the cost of data preprocessing as $c_V V_i^2$.

To model price and quality competition in the downstream market, we build on prior research \citep{tsay2000channel,banker1998quality} by assuming that aggregate consumer demand is linearly influenced by both price and product quality. We further assume that the quality of an AI-powered product corresponds to the quality of the fine-tuned model, denoted by $q_i$. Specifically, consumer demand for firm $i$ is given by:
\begin{equation*}
d_i = 1 - p_i - \theta_p (p_i - p_{j}) + q_i + \theta_q (q_i - q_{j}),
\end{equation*}
where $i,j\in\{1,2\}$ and $i \neq j$. Under this formulation, lowering a firm's price or improving its product quality increases its own demand while reducing that of its competitor, \textit{ceteris paribus}. The parameters $\theta_p > 0$ and $\theta_q > 0$ capture the intensity of price and quality competition in the consumer market, respectively. As firms generate revenue ($p_i d_i$) by selling AI-powered products to consumers, they simultaneously incur inference costs ($p_I d_i$) paid to the provider when consumers interact with the underlying model.

Accordingly, the profit of firm $i$ equals its sales revenue minus the costs of inference, fine-tuning, and data preprocessing. The resulting profit function for firm $i$ is given by:
\begin{align}
\label{firm profits}
\pi_i = (p_i - p_I) d_i - p_F V_i - c_V V_i^2.
\end{align}

\subsection{Model Timeline}
\begin{figure}
\FIGURE
{\includegraphics[width=0.7\textwidth]{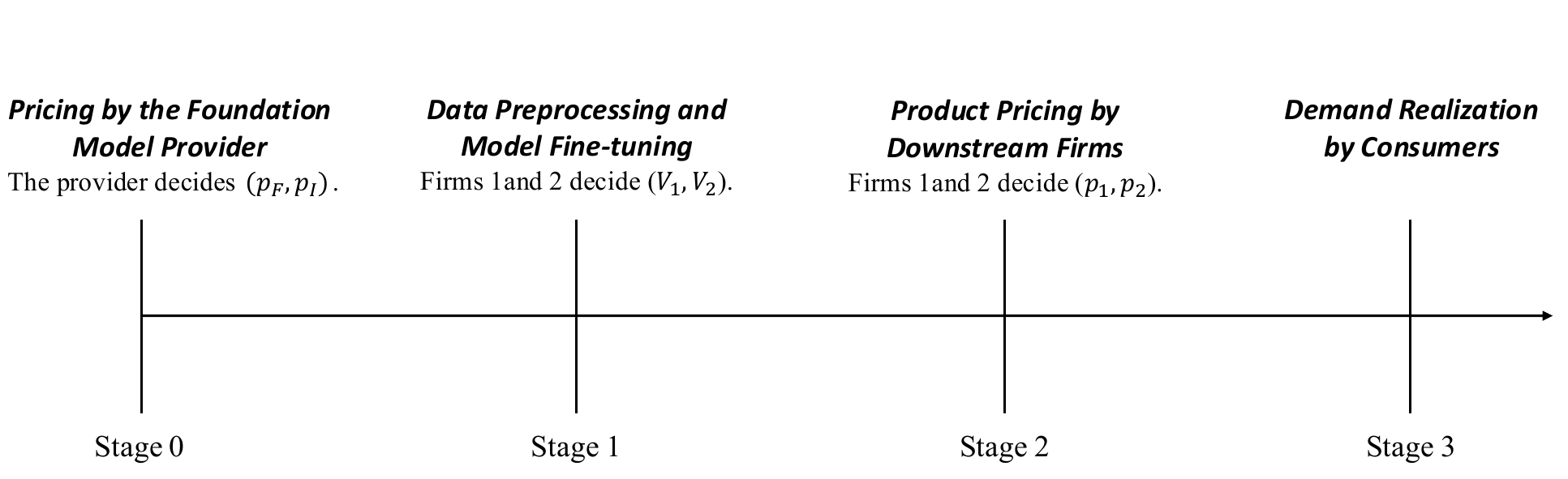}}
{Model Timeline \label{fig:time}}
{}
\end{figure}
The game unfolds in four stages, as illustrated in Figure \ref{fig:time}, which depicts the interactions among the key stakeholders, namely the foundation model provider, downstream firms, and consumers. In Stage 0, the provider sets the fine-tuning price ($p_F$) and the inference price ($p_I$), both of which are observable to the firms. In Stage 1, Firms 1 and 2 simultaneously choose their respective data volumes for fine-tuning ($V_i$, where $i\in\{1,2\}$), preprocess their domain-specific data, and fine-tune the foundation model accordingly. Once fine-tuning is complete, the firms develop AI-powered products based on the fine-tuned model, resulting in product quality denoted by $q_i$. In Stage 2, after observing each other’s product quality, the firms simultaneously set their product prices ($p_i$) and offer their AI-powered products to consumers. Finally, in Stage 3, consumer demand is realized. This sequential structure, in which quality decisions precede pricing, aligns with established models of quality and price competition in the literature \citep{banker1998quality, demirhan2005information}. The underlying rationale is that quality decisions are typically more difficult to modify than pricing decisions; thus, firms often determine product quality before setting prices \citep{banker1998quality, demirhan2005information}.

\section{Equilibrium Analysis}
In this section, we employ backward induction to derive the subgame-perfect Nash equilibrium (SPNE) for the players' actions at each stage of the game, highlighting our key findings. Following the sequence outlined in Figure \ref{fig:time}, we first determine the product prices and data volumes chosen by the downstream firms, based on the fine-tuning and inference pricing decisions made by the foundation model provider. We then analyze the provider's equilibrium choices with respect to these upstream decisions.

\subsection{Downstream Firms' Decisions}
In this subsection, we focus on the subgame starting at Stage 1, where the provider’s inference price ($p_I$) and fine-tuning price ($p_F$) are treated as given. Using backward induction, we begin our analysis at Stage 2, where downstream firms determine their product prices to maximize their respective profit functions, as specified in Equation~\eqref{firm profits}. In this stage, firms take as given the realized product qualities, along with the previously set prices for fine-tuning and inference. We then proceed to derive the best-response function of firm $i$ for Stage 2 as:
\begin{equation}
    p_i(V_i,V_{j},p_I) = \frac{p_I(1+\theta_p)(2+3\theta_p) + 2(1+\alpha+V_i+\theta_q (V_i-V_{j})) + \theta_p (3+ 3\alpha + 2V_i+V_{j}+\theta_q(V_i-V_{j}))}{(2+\theta_p)(2+3\theta_p)}, \label{pi at stage 0}
\end{equation}
 where $i \neq j \in \{1,2\}$.
 
By substituting the above expressions for prices into the firms' profit functions from Equation \eqref{firm profits}, we derive the firms' profits in Stage 1. In this stage, each firm selects its data volume ($V_i$) for fine-tuning to maximize its profit. Solving the first-order condition yields the SPNE data volume for firm $i$ (denoted as $V_i(p_F,p_I)$).\footnote{For a pure strategy equilibrium to exist, the reaction functions of the two firms must intersect exactly once. To ensure this condition, we impose the following assumption: $|\frac{\partial^2 \pi_1}{\partial {V_1}^2}|\geq |\frac{\partial^2 \pi_1}{\partial V_1 \partial V_2}|$ \citep{tirole1988theory,banker1998quality,demirhan2005information}. \label{footnote: PureStrategyCondition}} The corresponding product prices ($p_i(p_F,p_I)$) can be obtained by substituting $V_i(p_F,p_I)$ into Equation \eqref{pi at stage 0}. The following lemma summarizes the firms' equilibrium decisions in this subgame at Stage 1.
\begin{lemma} 
    [Equilibrium Decisions of the Firms in the Subgame] \label{lemma: firm's decisions}
    Given the inference price $p_I$ and fine-tuning price $p_F$ set by the foundation model provider, the equilibrium data volumes and product prices for Firm $i$ ($i\in\{1,2\}$) in the subgame at Stage 1 are as follows: 
\begin{equation}
    \label{SPNE Vi}
    {V_i}(p_I, p_F)=\frac{2(1+\theta_p)(1+\alpha-p_I)(2(1+\theta_q)+\theta_p(2+\theta_q))-(2+\theta_p)^2 (2+3\theta_p)p_F}{2c_V (2+\theta_p)^2 (2+3\theta_p)-2(1+\theta_p)(2(1+\theta_q)+\theta_p(2+\theta_q))},
\end{equation}
\begin{equation}
    \label{SPNE Pi}
    {p_i}(p_I, p_F) = 
    \frac{
    \substack{
       \textstyle 2c_V(2+\theta_p)(2+3\theta_p)(1+\alpha) + 2p_I(1+\theta_p)(c_V(2+\theta_p)(2+3\theta_p) \\
     \textstyle - \theta_p \theta_q - 2(1+\theta_p+\theta_q)) - (2+\theta_p)(2+3\theta_p)p_F}}
    { 2c_V(2+\theta_p)^2(2+3\theta_p) - 2(1+\theta_p)(2(1+\theta_q)+\theta_p(2+\theta_q))}.
\end{equation}
\end{lemma}

Accordingly, we can rewrite Firm~$i$’s profit function by substituting the above prices and data volumes into Equation~\eqref{firm profits}, yielding the following expression:
\begin{equation}
\label{SPNE pi}
\pi_i = \bigl(p_i(p_I,p_F) - p_I \bigr) \bigl( 1 + \alpha + V_i(p_I,p_F) - p_i(p_I,p_F)\bigr) - p_F V_i(p_I,p_F) - c_V V_i^2(p_I,p_F)
\end{equation}

As established in Lemma \ref{lemma: firm's decisions}, the equilibrium data volumes and product prices in the subgame are influenced by the intensity of the downstream competition, captured by the parameters $\theta_p$ (price competition) and $\theta_q$ (quality competition). Specifically, an increase in price competition leads to reductions in both data volumes and product prices (i.e., $\frac{\partial {V_i}(p_I,p_F)}{\partial \theta_p}<0$, $\frac{\partial {p_i}(p_I,p_F)}{\partial \theta_p}<0$). In contrast, an increase in quality competition results in higher data volumes and product prices (i.e., $\frac{\partial {V_i}(p_I,p_F)}{\partial \theta_q}>0$, $\frac{\partial {p_i}(p_I,p_F)}{\partial \theta_q}>0$). The underlying mechanism is as follows: As price competition intensifies, firms are prompted to lower their product prices (i.e., $\frac{\partial {p_i}(p_I,p_F)}{\partial \theta_p}<0$), thereby diminishing the marginal returns to quality improvements through increased consumer demand. Consequently, firms decrease the volume of data used for fine-tuning (i.e., $\frac{\partial {V_i}(p_I,p_F)}{\partial \theta_p}<0$). In contrast, heightened quality competition incentivizes firms to enhance product quality by increasing data volumes for fine-tuning (i.e., $\frac{\partial {V_i}(p_I,p_F)}{\partial \theta_q}>0$). This improvement in quality enables firms to command higher prices (i.e., $\frac{\partial {p_i}(p_I,p_F)}{\partial \theta_q}>0$), since consumers are willing to pay a premium for the superior quality of the product resulting from more extensive data use. We now proceed to examine the provider's decision-making in Stage 0.


\subsection{Foundation Model Provider's Decisions}
In this subsection, we analyze the the foundation model provider's equilibrium pricing decisions. To this end, we substitute SPNE prices and data volumes into the provider's profit function, as specified in Equation \eqref{foundation model provider profits}. This allows us to rewrite the provider's profit function as:
\begin{equation}
    \pi_M = \sum_{i=1}^{2} \left(p_I d_i(p_I, p_F) + p_F V_i(p_I, p_F)- c_F V_i^2(p_I, p_F)\right), \label{pi_M at stage 0}   
\end{equation}
where $d_i(p_I,p_F)=1+\alpha+V_i(p_I,p_F)-p_i(p_I,p_F)$. The provider chooses the inference price ($p_I$) and the fine-tuning price ($p_F$) to maximize profit, such that $\pi_M^*=\max_{p_I,p_F} \pi_M$.

By solving the corresponding first-order conditions, we obtain the equilibrium inference and fine-tuning prices, denoted by $p_I^*$ and $p_F^*$, respectively.\footnote{We focus on non-negative interior solutions for $(p_I^*, p_F^*, p_i^*, V_i^*)$ to ensure a non-trivial equilibrium. In addition to the pure-strategy equilibrium conditions specified in Footnote \ref{footnote: PureStrategyCondition}, we impose the technical constraints $c_F \geq \underline{c}_F$ and $c_V \geq \underline{c}_V$, with expressions for $\underline{c}_F$ and $\underline{c}_V$ provided in the Appendix.} These results are summarized in the following lemma.

\begin{lemma}
[Equilibrium Pricing Decisions of the Provider] \label{lemma: equilibrium}
In equilibrium, the foundation model provider sets the inference and fine-tuning prices as follows:\\
\begin{equation*}
p_I^* = \frac{\substack{\textstyle 2(1+\alpha)(c_F (2+\theta_p)^3(2+3\theta_p)^2+2c_V (2+\theta_p)^3(2+3\theta_p)^2 \\\textstyle -(1+\theta_p)(2+2\theta_p+2\theta_q+\theta_p\theta_q)(8+12\theta_p+3\theta_p^2+2(2+\theta_p)\theta_q))}}{4c_F (2+\theta_p)^3(2+3\theta_p)^2+8c_V (2+\theta_p)^3(2+3\theta_p)^2-(1+\theta_p)(8+12\theta_p+3\theta_p^2+2(2+\theta_p)\theta_q)^2},
\end{equation*}
\begin{equation*}
p_F^*=\frac{2(1+\theta_p)(2+\theta_p)(2+3\theta_p)(1+\alpha)(4c_F + 4c_F \theta_p-4c_V \theta_p -3c_V \theta_p^2+2(c_F+c_V)(2+\theta_p)\theta_q)}{4c_F(2+\theta_p)^3(2+3\theta_p)^2+8c_V(2+\theta_p)^3(2+3\theta_p)^2-(1+\theta_p)(8+12\theta_p+3\theta_p^2+2(2+\theta_p)\theta_q)^2}.
\end{equation*}
\end{lemma}

Accordingly, the equilibrium data volumes and product prices, denoted by $V_i^*$ and $p_i^*$, are obtained by substituting $p_I^*$ and $p_F^*$ into Equations~\eqref{SPNE Vi} and \eqref{SPNE Pi}, respectively.
As shown in Lemma~\ref{lemma: equilibrium}, the equilibrium prices of the provider depend on the parameters of the downstream market $(\theta_p,\theta_q)$. Therefore, we examine how the intensity of downstream price and quality competition affects the provider's optimal pricing decisions, as summarized in the following lemmas.

\begin{lemma} [Effects of Downstream Competition on the Fine-Tuning Price]
\label{lemma: competition affect fine-tuning price}  
\quad
\begin{enumerate}[label=\alph{enumi}., ref=\alph{enumi}]
		\item The fine-tuning price decreases with the intensity of downstream price competition (i.e.,
        $\frac{d p_F^*}{d \theta_p}<0$). \label{pF with theta_p}
		\item The fine-tuning price increases with the intensity of downstream quality competition (i.e.,
        $\frac{d p_F^*}{d \theta_q}>0$). \label{pF with theta_q}
	\end{enumerate}
\end{lemma}

Lemma~\ref{lemma: competition affect fine-tuning price} demonstrates that the fine-tuning price decreases with downstream price competition ($\theta_p$) and increases with downstream quality competition ($\theta_q$). As price competition intensifies, firms reduce the data volumes used for fine-tuning, as explained following Lemma~\ref{lemma: firm's decisions}. This reduction lowers not only fine-tuning revenue but also inference revenue (due to the resulting degradation in product quality). The provider therefore has an incentive to lower the fine-tuning price to stimulate data volume, leading to $\frac{d p_F^*}{d \theta_p}<0$. In contrast, as quality competition intensifies, firms increase data volume, as discussed after Lemma~\ref{lemma: firm's decisions}, prompting the provider to raise the fine-tuning price. 

The next lemma examines how downstream price and quality competition affect the inference price.

\begin{lemma} [Effects of Downstream Competition on Inference Price]
\label{lemma: competition affect inference price}
\quad
\begin{enumerate}[label=\alph{enumi}., ref=\alph{enumi}]
    \item The inference price increases with the intensity of downstream price competition ($\frac{d p_I^*}{d \theta_p}>0$) when $c_F+2 c_V>c_1$ and decreases otherwise. 
    \item The inference price decreases with the intensity of downstream quality competition ($\frac{d p_I^*}{d \theta_q}<0$) when $c_F+2 c_V>c_2$ and increases otherwise. \label{pI with theta_q}
\end{enumerate}

For brevity, the expressions for $c_1$ and $c_2$ are provided in the Online Appendix.
\end{lemma}

In contrast to the fine-tuning price, both downstream price and quality competition have non-monotonic effects on the inference price, as indicated by Lemma~\ref{lemma: competition affect inference price}. The outcome arises because price competition influences inference prices in two distinct ways. First, intensified price competition reduces product prices, which, in turn, increases inference demand, prompting the provider to raise the inference price. Second, lower product prices, resulting from heightened price competition, reduce the marginal benefits of product quality, leading firms to invest less in quality improvement. 
This is achieved by decreasing the data volume used for fine-tuning. The reduction in data volume diminishes the provider's revenue, prompting a decrease in the inference price, which increases the marginal benefit of quality improvement and thereby stimulates data volume. 
However, the magnitude of this effect depends on the firms' marginal cost of data volume. Specifically, it can be shown that when $c_F$ or $c_V$ is relatively high (i.e., $c_F>c_1-2c_V$ or $c_V>(c_1-c_F)/2$), the marginal cost of data volume is also high. In such cases, increasing the data volume by reducing the inference price becomes more difficult. Consequently, the provider may not reduce the inference price substantially, leading to an overall increase in the inference price. Conversely, when the marginal cost of data volume is relatively low, the second effect dominates. A similar reasoning applies to the impact of increased quality competition on the inference price.

\section{Consumer Surplus and Policy Implications}
\label{sec: Consumer Surplus and Policy Implications}
With the growing use of AI-enabled products, consumer welfare faces significant risks, such as the difficulty of assessing product quality \citep{AIProductQualityHardToAssess2}.
In light of these concerns, policymakers are increasingly focused on preserving consumer welfare in the context of AI \citep{ConsumerProtectionEraOfAI}.
However, the complex interdependencies between upstream providers and downstream markets, as discussed in Section 4, highlight the regulatory challenges inherent to the AI supply chain \citep{RegulationChallenging}. 
This section addresses these issues by evaluating the effectiveness of existing regulatory frameworks and providing guidance for selecting appropriate policy instruments. We focus on two prominent approaches: promoting downstream market competition and providing compute subsidies \citep{CompetitionBureauCanada, BeijingSubsidy}. Our analysis proceeds in two parts: first, we assess the impact of pro-competitive policies on consumer surplus; second, we evaluate the effectiveness of compute subsidy policies. We conclude by discussing the conditions under which each policy instrument most effectively enhances consumer surplus.

\subsection{Pro-Competitive Policies in Downstream Markets} 
\label{section: competition}
For policymakers, pro-competitive regulation has long been a key strategy to enhance consumer welfare, primarily by lowering prices and improving product quality \citep{tucker2024does}. 
To reduce prices, a key component of pro-competitive policies is promoting price competition in the consumer market (hereafter referred to as the \textit{pro-price-competitive policy}). This can be achieved through measures such as enhancing price transparency or facilitating product price comparisons. 
One prominent instance is the Federal Trade Commission's (FTC) Rule on Unfair or Deceptive Fees. This rule requires companies to disclose the total price in advance, avoiding hidden or misleading fees \citep{FTCRules}.
In a recent case, the FTC accused Ticketmaster, a leading online ticketing platform, of violating this rule by failing to provide complete price transparency. As a result, the company was required to adjust its pricing system to ensure that consumers could see the total cost before making a purchase \citep{Ticketmaster}. Similarly, the U.S. Federal Communications Commission (FCC) has implemented regulations requiring Internet service providers to display their base monthly prices on standardized labels.  This rule aims to improve consumer decision-making by making it easier for individuals to compare prices across providers, thus enhancing market transparency and fostering more competitive pricing \citep{FCCRules}. 

Another key component of pro-competitive policies is fostering quality competition in consumer markets (hereafter referred to as the \textit{pro-quality-competitive policy}), often by increasing transparency about product quality. A notable example is the FTC's \textit{Trade Regulation Rule on the Use of Consumer Reviews and Testimonials}, which aims to prevent deceptive practices such as fake reviews and the suppression of negative feedback \citep{FTCRules2}. In one case, the FTC alleged that Fashion Nova misrepresented its product reputation by concealing customer reviews with ratings below four stars, thus misleading consumers about the true quality of its products \citep{FashionNova}.

In the context of the AI supply chain, policymakers are increasingly advocating for the development of similar pro-competitive policies \citep{CompetitionBureauCanada}. 
The FTC, for example, has issued guidance that prohibits companies from exaggerating or misrepresenting the capabilities of their AI-based products \citep{FTCAIGuideline}. 
Companies found guilty of overstating the capabilities of their AI products, such as DoNotPay, have faced FTC complaints and proposed settlements for deceptive claims \citep{FTCPunishDeceptiveAIClaims}.
Similarly, the European Union’s \textit{Consumer Rights Act} requires businesses to disclose the total price of digital content or services, including AI-based products, and prohibits hidden fees \citep{ConsumerRightAct}. Together, these policies promote greater transparency regarding both price and quality, thereby fostering more intense competition in the downstream segment of the AI supply chain.
\footnote{Studies show that greater market transparency often leads to increased competition, ultimately benefiting consumers \citep{schultz2009transparency}.}

Recent research and policy analyses highlight that increased competition in product markets can benefit consumers by driving down prices or encouraging quality improvements \citep{BenefitOfCompetition, tyagi1999effects, wu2022competition}. However, the existing literature also points to potential downsides, suggesting that intensified price competition may diminish firms' incentives to invest in product quality \citep{tsay2000channel}, ultimately leading to lower product quality and reduced consumer surplus. A similar concern arises in quality competition, where research indicates that aggressive competitive pressures may lead to a decrease in product quality \citep{armstrong2009inattentive}. In the context of the AI supply chain, it remains unclear which of these opposing effects dominates, raising questions about whether pro-competitive policies in downstream markets ultimately enhance or reduce consumer surplus. 

To address this issue, we first define consumer surplus, denoted by $CS$, within the framework of our model. Following the literature \citep{singh1984price, wang2024economics}, we consider a representative consumer and derive $CS$ from the corresponding utility function (i.e., $U(d_i, d_{j})=\sum_{i=1}^2 \left(\mu_i d_i-\frac{\beta_i d_i^2+\gamma d_i d_{j}}{2}-p_i d_i\right)$, where the first term represents the direct benefit of consuming the products, the second term accounts for the diminishing returns, and the third term reflects the payment made by the consumer. The resulting expression for $CS$ and its derivation are provided in the Online Appendix.
In equilibrium, this expression simplifies to:
\begin{equation} \label{eq:equilibriumCS}
CS = \sum_{i=1}^2 \frac{1}{2} (1+\alpha+V_i^*-p_i^*)^2,
\end{equation}
which indicates that consumer surplus is jointly determined by product quality ($q_i^*=\alpha+V_i^*$) and prices ($p_i^*$). 

Based on Equation \eqref{eq:equilibriumCS}, we analyze the effects of pro-price-competitive and pro-quality-competitive policies in the downstream market on consumer surplus. The results are summarized in the following proposition. \looseness=-1

\begin{proposition}
[Impact of Pro-competitive Policies on Consumer Surplus]
\label{prop: competition}
\quad
\begin{enumerate}[label=\alph{enumi}., ref=\alph{enumi}]
	\item As price competition in downstream markets intensifies, \label{price competition reduce consumer surplus}
	consumer surplus decreases if $c_F+2 c_V<\tilde{c}$, and increases otherwise.
    \item As quality competition in downstream markets intensifies, consumer surplus increases. \label{quality competition increases consumer surplus}
\end{enumerate} 
The explicit expression for \( \tilde{c} \) is algebraically complex and is thus provided in the Online Appendix.
\end{proposition}

As shown in Proposition \ref{prop: competition}\ref{price competition reduce consumer surplus}, while pro-competitive policies may intensify price competition and reduce product prices, consumer surplus can paradoxically decline when compute costs (i.e., $c_F$) or data preprocessing costs (i.e., $c_V$) are relatively low. This counter-intuitive result can be explained as follows.

As price competition intensifies, firms are incentivized to lower their product prices, which benefits consumers (i.e., $\frac{d p_i^*}{d \theta_p}<0$). However, the impact on product quality is more nuanced, as it can move in either direction due to the co-creation dynamics inherent in the AI supply chain. The direct effect of increased price competition is a reduction in product quality (i.e., $\frac{\partial V_i(p_I,p_F)}{\partial \theta_p}\big|_{p_I^*,p_F^*}<0$): Lower prices reduce the marginal benefit of offering higher quality, thus diminishing firms' incentives to invest in product quality. 

In contrast, the indirect effect operates through the provider's pricing response (i.e., $\bigl(\frac{\partial V_i(p_I,p_F)}{\partial p_I}\frac{d p_I^*}{d \theta_p}+\frac{\partial V_i(p_I,p_F)}{\partial p_F}\frac{d p_F^*}{d \theta_p}\bigr)|_{p_I^*,p_F^*}$). 
Specifically, as price competition intensifies, the provider lowers the fine-tuning price (i.e., $\frac{d p_F^*}{d \theta_p}<0$), while the inference price may rise or fall (i.e., $\frac{d p_I^*}{d \theta_p}$ can take either sign), as shown in Lemmas~\ref{lemma: competition affect fine-tuning price} and \ref{lemma: competition affect inference price}. A higher inference price reduces the marginal benefit of product quality, leading to a decrease in data volumes ($\frac{\partial V_i(p_I,p_F)}{\partial p_I}\big|_{p_I^*,p_F^*}<0$), while a lower fine-tuning price reduces the marginal cost of data volume, thus increasing data volumes ($\frac{\partial V_i(p_I,p_F)}{\partial p_F}\big|_{p_I^*,p_F^*}<0$). Even if the inference price increases and suppresses data volumes, the effect of the reduction in the fine-tuning price dominates, resulting in an overall increase in product quality ($\bigl(\frac{\partial V_i(p_I,p_F)}{\partial p_I}\frac{d p_I^*}{d \theta_p}+\frac{\partial V_i(p_I,p_F)}{\partial p_F}\frac{d p_F^*}{d \theta_p}\bigr)\big|_{p_I^*,p_F^*}>0$). Taken together, intensified price competition indirectly encourages higher product quality by enhancing the provider's co-creation incentives by lowering the fine-tuning price (and sometimes the inference price). 
However, as explained in the following, when the coefficients for the data preprocessing costs and the compute costs (i.e., $c_V$ and $c_F$) are relatively low, this indirect effect is outweighed by the direct effect. In this case, the net result is a decline in product quality, thereby reducing consumer surplus.

When $c_V$ increases, the firms' marginal cost of quality improvement rises, which reduces the direct effect on product quality (i.e., $
\frac{d}{d c_V} \big( \frac{\partial V_i(p_I, p_F)}{\partial \theta_p} \big|_{ p_I^*,p_F^*} \big) > 0$). 
A similar effect arises when $c_F$ increases: Increased compute costs lead to a higher fine-tuning price, which in turn increases the marginal cost of quality improvement, further diminishing the direct effect (i.e., $
\frac{d}{d c_F} \big( \frac{\partial V_i(p_I, p_F)}{\partial \theta_p} \big|_{ p_I^*,p_F^*} \big) > 0$). 
Thus, when $c_V$ or $c_F$ is relatively high, the negative direct effect of intensified price competition on product quality is dominated, leading to increased consumer surplus as $\theta_p$ increases.
In contrast, when both $c_V$ and $c_F$ are relatively low (i.e., $c_F+2c_V<\tilde{c}$), the direct effect on product quality dominates, leading to a decline in product quality and, consequently, consumer surplus as $\theta_p$ increases.

Proposition \ref{prop: competition}\ref{quality competition increases consumer surplus} further demonstrates that consumers always benefit as quality competition in the downstream market intensifies. At first glance, this may seem intuitive, as increased quality competition encourages firms to enhance product quality (i.e., $\frac{d V_i^*}{d \theta_q}>0$). However, our analysis uncovers an additional, more nuanced mechanism: In certain cases, intensified quality competition also leads to lower product prices (i.e., $\frac{d p_i^*}{d \theta_q}<0$ if $c_F+2c_V>\frac{(1+\theta_p)(3\theta_p^2+4(2+\theta_q)+2\theta_p(6+\theta_q))^2}{4(2+\theta_p)(2+3\theta_p)(\theta_p^2+2(1+\theta_p)(2+\theta_p)\theta_q)}$), which in turn benefits consumers. This outcome arises from indirect effects driven by the strategic pricing behavior of the foundation model provider within the AI supply chain. Specifically, as quality competition intensifies, the provider may lower the inference price (as established in Lemma~\ref{lemma: competition affect inference price}), which in turn incentivizes firms to reduce their product prices. 
Thus, the increase in consumer surplus from intensified quality competition is not solely driven by higher product quality; it can also stem from lower product prices. This challenges the conventional view that higher-quality products always command higher prices, as suggested in the literature (e.g., \citealp{tirole1988theory, sharma2021entry}). \looseness=-1


\subsection{Compute Subsidies}
\label{section: computing power vouchers}
In addition to pro-competitive policies, subsidizing access to computing power (hereafter referred to as \textit{compute subsidies}) is a widely used regulatory tool employed by policymakers. By reducing compute costs, these subsidies incentivize firms to leverage more data for model fine-tuning, ultimately improving product quality and increasing consumer surplus \citep{sastry2024computing}.
In China, the government has implemented this approach within the AI supply chain.
For example, the Beijing municipal government allocates annual compute subsidies totaling 100 million RMB to firms involved in large-scale AI model services. 
These subsidies can offset up to 30\% of computing expenses, including those associated with fine-tuning \citep{BeijingSubsidy}. 
Similarly, the Shenzhen municipal government has introduced an annual compute subsidy program worth 500 million RMB, covering up to 50\% of compute costs. 
Comparable initiatives have also been launched by the Shanghai and Chengdu municipal governments \citep{BeijingSubsidy}.

To analyze the impact of this policy on consumer surplus, we consider a scenario in which policymakers offer compute subsidies that reduce compute costs by a proportion $x$. 
Here, $x$ denotes the subsidy rate, defined as the fraction of compute costs covered by policymakers.
Under this policy, the foundation model provider incurs fine-tuning costs of $c_F(1-x)V_1^2+c_F(1-x)V_2^2$, rather than the full cost $c_FV_1^2+c_FV_2^2$. Let $V_1'$ and $V_2'$ denote the resulting equilibrium data volumes, and $CS'$ represent the corresponding consumer surplus under the compute subsidy. The total subsidy expenditure under this policy is given by $x c_F V_1'^2 + x c_F V_2'^2$. We then compare the consumer surplus with and without the policy and present the results in the following lemma.
\begin{lemma}
    [Impact of Compute Subsidies on Consumer Surplus]
\label{lemma: subsidy increase consumer surplus}
When compute subsidies are provided, consumer surplus increases compared to the scenario without them (i.e., \( \forall x \in [0,1],\ CS^{\prime}(x) > CS(x) \)).
\end{lemma}

Lemma~\ref{lemma: subsidy increase consumer surplus} shows that, although compute subsidies are granted to the foundation model provider, they ultimately benefit consumers. This can be explained as follows: When policymakers subsidize compute costs, the foundation model provider faces a lower marginal compute cost, which leads to a reduction in the fine-tuning price. The lower fine-tuning price encourages firms to increase their data volumes for fine-tuning, thereby improving product quality and enhancing consumer surplus. 

Although compute subsidies increase consumer surplus, it is essential to assess whether the policy is cost-effective, particularly given the substantial public funds involved. To do so, we adopt the framework from the literature \citep{anderson2022bike} and define \textit{cost-effectiveness} ($CE$) as the net benefits of compute subsidies, which are measured by the difference between the incremental consumer surplus obtained from compute subsidies and the subsidy expenditure, and can be expressed as $CS'-CS-x c_F V_1'^2-x c_F V_2'^2$. A positive value indicates that the policy is cost-effective in enhancing consumer surplus. We formally analyze this effectiveness metric and present the results in the following proposition.

\begin{proposition}
[Cost-Effectiveness of Compute Subsidies]
\label{prop: subsidy}
Compute subsidies are cost-effective only when $c_F+2 c_V<\hat{c}$.
The explicit expression for $\hat{c}$ is algebraically complex and is thus provided in the Online Appendix.
\end{proposition}

Although consumer surplus increases with compute subsidies (as shown in Lemma~\ref{lemma: subsidy increase consumer surplus}), Proposition~\ref{prop: subsidy} indicates that such subsidies are cost-ineffective in enhancing consumer surplus when compute and data preprocessing costs are relatively high. This counter-intuitive result can be explained as follows: While compute subsidies improve product quality and thus enhance consumer surplus (as discussed following Lemma~\ref{lemma: subsidy increase consumer surplus}), they also incur substantial costs. When these costs outweigh the increase in consumer surplus, the subsidies become cost-ineffective. 
When the coefficient of data preprocessing costs ($c_V$) is relatively high, firms face substantial marginal costs in expanding data volumes for fine-tuning, limiting their ability to improve product quality even with subsidized compute costs.
As a result, the increase in consumer surplus is less than the subsidy expenditure. 
Similarly, when the coefficient of compute costs ($c_F$) is high, the provider still faces significant marginal costs despite the subsidy and is thus incentivized to maintain relatively high fine-tuning prices. This results in high marginal costs for firms that attempt to improve product quality, thereby limiting the policy's cost-effectiveness. Therefore, when $c_F$ or $c_V$ is relatively high, compute subsidies are cost-ineffective in improving consumer surplus.

As shown in Proposition~\ref{prop: subsidy}, compute subsidies can be cost-effective when compute costs and data preprocessing costs are relatively low. Specifically, when $c_F+2c_V<\hat{c}$, there exists a subsidy rate (i.e., $x$) such that the net benefits of compute subsidies are positive (i.e., $CE>0$). 
This indicates that whether these subsidies are cost-effective also depends on the choice of the subsidy rate.
Thus, we examine the relationship between cost-effectiveness and the subsidy rate for computing power under the condition ($c_F+2c_V<\hat{c}$), and summarize our findings in the following corollary.


\begin{corollary}
[Relationship between Cost-Effectiveness and Subsidy Rate for Compute Costs]
\label{coro: subsidy}
As the subsidy rate for compute costs (i.e., $x$) increases, cost-effectiveness improves if $x<x'$, and decreases if $x>x'$. Additionally, if $x>x''$ (with $x'' > x'$), compute subsidies become cost-ineffective; otherwise, they become cost-effective. 

For brevity, the explicit expressions for $x'$ and $x''$ are provided in the Online Appendix.
\end{corollary}

Corollary~\ref{coro: subsidy} demonstrates that the net benefits of compute subsidies do not increase monotonically with the subsidy rate $x$. Instead, the net gain in consumer surplus relative to the subsidy costs initially rises and then falls as $x$ increases, with cost-effectiveness peaking at $x'$. This non-monotonic pattern can be explained by the fine-tuning behavior of downstream firms: At low subsidy rates, increasing $x$ encourages firms to expand data volumes for fine-tuning, which improves product quality, increases consumer surplus, and enhances cost-effectiveness. This explains the initial upward trend for values of $x$ below $x'$. However, once the subsidy rate exceeds this threshold (i.e., $x > x'$), the provider responds by lowering the fine-tuning price excessively, which incentivizes firms to engage in more aggressive fine-tuning. Although this improves product quality and further increases consumer surplus, it also substantially increases data volume, thereby driving up subsidy costs ($x c_F V_1'^2 + x c_F V_2'^2$). As a result, the gains in consumer surplus are eventually outweighed by the increase in subsidies when $x > x'$, thereby reducing the cost-effectiveness of compute subsidies. If the subsidy rate $x$ exceeds $x''$, the cost-effectiveness may even turn negative. Therefore, under the condition $c_F+2 c_V<\hat{c}$, compute subsidies remain cost-effective only if the subsidy rate is not too high (i.e., $x<x''$). \looseness=-1

Proposition~\ref{prop: subsidy} offers valuable regulatory insights for policymakers. The findings suggest that compute subsidies must be implemented with caution, as their effectiveness in enhancing consumer surplus depends heavily on the cost environment. Specifically, when data preprocessing costs or compute costs are relatively high, such subsidies yield limited benefits to consumer surplus and may even fail to cover the subsidy expenditures. 
Additionally, to ensure the effectiveness of compute subsidies, it is also crucial to carefully calibrate the subsidy rate. If the rate is set too high, it could encourage downstream firms to overinvest in fine-tuning, leading to disproportionately high subsidy expenditures and diminishing, or even negative, returns in terms of net consumer surplus gains.


\subsection{Comparing Policies}
\begin{figure}
\FIGURE
{\includegraphics[width=1\textwidth]{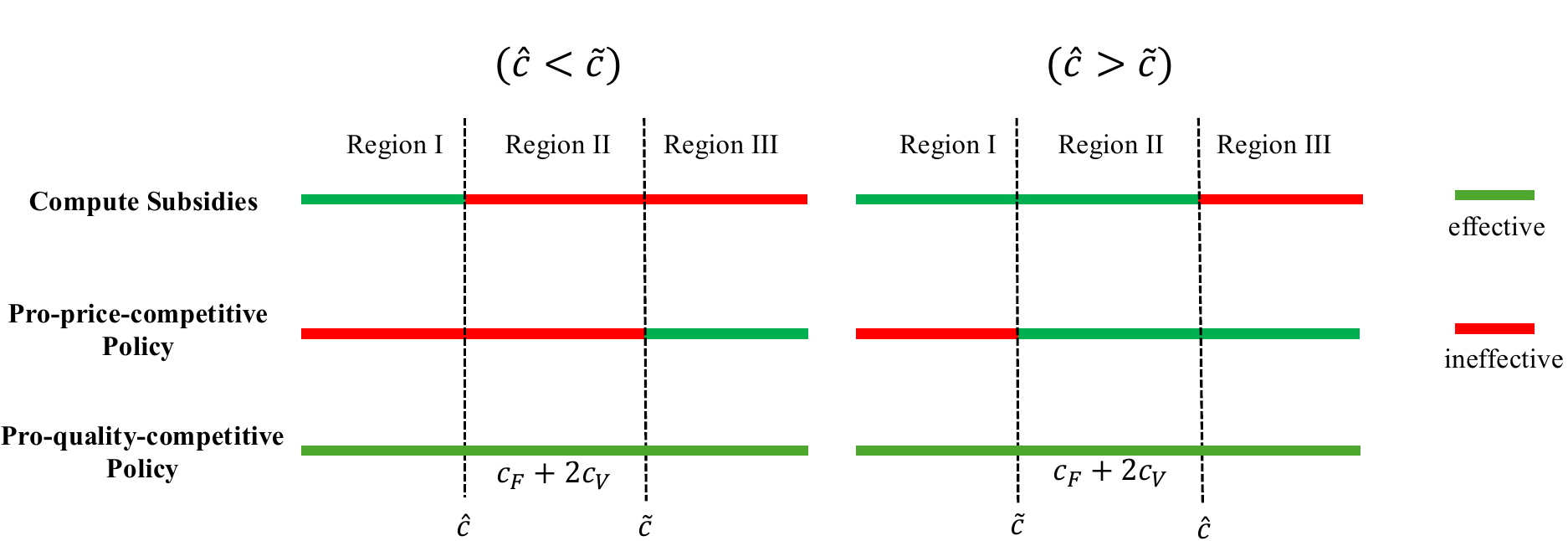}}
{Policy Effectiveness under Different Cost Regions
\label{fig: policy consideration sets}}
{}
\end{figure}

Building on the previous analysis, this subsection compares the effectiveness of each policy in enhancing consumer surplus across different cost regions.
Unlike compute subsidies, which impose a direct cost on the policymaker, the pro-competitive policy incurs no government expenditure. Thus, the effectiveness of the pro-competitive policy is assessed solely by its ability to increase consumer surplus. In contrast, compute subsidies are deemed effective only when the increase in consumer surplus exceeds the costs incurred in providing the subsidy.
Figure~\ref{fig: policy consideration sets} illustrates the effective and ineffective policy regimes for compute subsidies, the pro-price-competitive policy, and the pro-quality-competitive policy. 
We divide the cost space into three distinct regions. 
When the compute and data preprocessing costs are relatively low ($c_F + 2c_V < \min\{\tilde{c}, \hat{c}\}$), the cost condition falls into Region~I. 
When the costs are moderate ($\min\{\tilde{c}, \hat{c}\} < c_F + 2c_V < \max\{\tilde{c}, \hat{c}\}$), this corresponds to Region~II. 
Finally, when the costs are relatively high ($c_F + 2c_V > \max\{\tilde{c}, \hat{c}\}$), this scenario falls into Region~III.
The regions are depicted differently depending on the relationship between $\hat{c}$ and $\tilde{c}$: The left panel shows the case where $\hat{c}<\tilde{c}$, while the right panel illustrates the case where $\hat{c}>\tilde{c}$.

As shown in Figure~\ref{fig: policy consideration sets}, the results of Proposition~\ref{prop: competition} and~\ref{prop: subsidy} reveal a complementarity between the pro-price-competitive policy and compute subsidies: When one is ineffective, the other is effective. 
Specifically, in Region~I, where the costs are relatively low ($c_F + 2c_V < \min\{\tilde{c}, \hat{c}\}$), the pro-price-competitive policy does not improve consumer surplus, whereas compute subsidies are cost-effective in doing so. 
Conversely, in Region~III, where the costs are relatively high, compute subsidies are cost-ineffective, while the pro-price-competitive policy becomes effective.

Our research provides important policy implications for policymakers seeking to enhance consumer surplus in the AI supply chain, offering structured policy recommendations under varying cost conditions. As illustrated by the green lines in Region I of Figure~\ref{fig: policy consideration sets}, where the compute and data preprocessing costs are relatively low ($0<c_F+2c_V<\min\{\hat c,\tilde c\}$), policymakers should focus solely on promoting competition in downstream markets, whether in quality or price. In this case, as indicated by the red line in Region I, compute subsidies are not cost-effective because firms are already engaging in aggressive fine-tuning, leading to excessive public expenditure on subsidies with limited additional benefit to consumer surplus. When these costs are relatively high ($c_F+2c_V>\max\{\hat c,\tilde c\}$), as illustrated by the green lines in Region III of Figure~\ref{fig: policy consideration sets}, policymakers should instead consider compute subsidies and pro-quality-competition policies. In this case, promoting price competition would reduce consumer surplus by discouraging firms from investing in data for fine-tuning, which in turn lowers product quality. When these costs fall within an intermediate range ($\min\{\hat c,\tilde c\}<c_F+2c_V<\max\{\hat c,\tilde c\}$), policymakers should consider adopting pro-quality-competitive policy if $\hat c<\tilde c$, as illustrated by the green line in Region II of the left panel; otherwise ($\hat c>\tilde c$), they should consider all three policies, as illustrated by the green lines in Region II of the right panel.

\section{Effect of Different Regulatory Policies on Profits}
\label{section: profits of foundation model provider and firms}
In addition to consumer surplus, it is crucial to understand how pro-competitive policies and compute subsidies impact the profitability of both foundation model providers and downstream firms, as the success of any regulatory policy ultimately hinges on its acceptance by industry participants. 
A policy that not only enhances consumer surplus but also improves the profitability of both providers and downstream firms is more likely to gain traction, thereby creating a ``win–win–win'' scenario in which all stakeholders benefit.
In this section, we begin by conducting a comparative statics analysis to assess how policies that promote competition in downstream markets impact the profits of both the foundation model provider and downstream firms, followed by an analysis of the effects of compute subsidies.

Understanding the effect of downstream competition on the profitability of both upstream and downstream firms remains a central concern in the study of industrial organization \citep{basak2023competition}. The conventional view holds that increased competition among downstream firms reduces their profitability while enhancing the profits of upstream firms \citep{basak2023competition,tyagi1999effects}. This perspective is based on the concept of double marginalization, which suggests that the increased competition among downstream firms weakens their pricing power, resulting in lower product prices (i.e., reduced double marginalization) and, in turn, increased demand for the product \citep{tyagi1999effects}. As a result, upstream firms are expected to see higher profits from the increase in sales to downstream firms, whereas downstream firms face shrinking margins and lower profitability. However, our analysis challenges this conventional wisdom, revealing that these typical outcomes do not necessarily hold in our setting. We summarize our findings in the following proposition and corollary.

\begin{proposition}
[Impact of Pro-competitive Policies on the Foundation Model Provider’s Profits]
\label{prop: provider profits with competition}
\indent 
\vspace{-\baselineskip}
\begin{enumerate}[label=\alph{enumi}., ref=\alph{enumi}]
\item As price competition intensifies, the profits of the foundation model provider decrease if $\theta_q>\tilde{\theta}_q$, and increase otherwise.
\item As quality competition intensifies, the foundation model provider's profits always increase. \label{item: quality competition increases provider profits}
\end{enumerate}
For brevity, the expression for \( \tilde{\theta}_q \) is provided in the Online Appendix.
\end{proposition}

This result highlights that intensified price competition in the downstream market can, under certain conditions, reduce the foundation model provider's profits. To better understand how downstream price competition affects the provider's profitability, we apply the chain rule to analyze $\frac{d \pi_M}{d \theta_p}\big|_{p_I^*, p_F^*}$ and present the simplified expression below, with the complete derivation available in the Online Appendix:
\begin{equation}
\label{simplified dpimdthetap}
\frac{d \pi_M}{d \theta_p}\Bigg|_{p_I^*, p_F^*}=\sum_{i=1}^2 \biggl(
\frac{\partial {\pi}_M}{\partial p_i}\frac{\partial p_i(p_I,p_F)}{\partial \theta_p}
+\frac{\partial {\pi}_M}{\partial V_i} 
\frac{\partial V_i(p_I,p_F)}{\partial \theta_p}\biggr)\Bigg|_{p_I^*, p_F^*},
\end{equation}
where the expression for ${\pi}_M$ is provided in Equation~\eqref{pi_M at stage 0}, with $\frac{\partial {\pi}_M}{\partial p_i}\big|_{p_I^*, p_F^*}=-p_I^*$
 and $\frac{\partial {\pi}_M}{\partial V_i}\big |_{p_I^*, p_F^*}=p_I^*+p_F^*-2c_F V_i^*$.


Equation~\eqref{simplified dpimdthetap} demonstrates that intensified price competition influences provider profits by shaping downstream firms’ decisions regarding both pricing and data usage for finetuning. On the one hand, consistent with conventional wisdom (i.e., reduced double marginalization), increased price competition prompts firms to lower their product prices ($\frac{\partial p_i(p_I,p_F)}{\partial \theta_p}\big |_{p_I^*, p_F^*}<0$), which stimulates product demand and, in turn, increases the provider's inference revenue ($\frac{\partial {\pi}_M}{\partial p_i}\frac{\partial p_i(p_I,p_F)}{\partial \theta_p}\big |_{p_I^*, p_F^*}=-p_I^* \frac{\partial p_i(p_I,p_F)}{\partial \theta_p}\big |_{p_I^*, p_F^*}>0$). On the other hand, intensified price competition encourages downstream firms to reduce their data usage for fine-tuning ($\frac{\partial V_i(p_I,p_F)}{\partial \theta_p}\big |_{p_I^*, p_F^*}<0$), as discussed following Lemma~\ref{lemma: firm's decisions}. This decline in fine-tuning demand not only reduces fine-tuning revenue ($ p_F^* \frac{\partial V_i(p_I,p_F)}{\partial \theta_p}\big |_{p_I^*, p_F^*}<0$), but also degrades product quality, which in turn reduces inference revenue ($p_I^* \frac{\partial V_i(p_I,p_F)}{\partial \theta_p}\big |_{p_I^*, p_F^*}<0$). Additionally, reduced data usage lowers the compute costs associated with fine-tuning ($2c_F V_i^* \frac{\partial V_i(p_I,p_F)}{\partial \theta_p}\big |_{p_I^*, p_F^*}<0$). However, despite the savings in compute costs, the overall effect of reduced data usage is a net decrease in provider profits ($\frac{\partial {\pi}_M}{\partial V_i}\frac{\partial V_i(p_I,p_F)}{\partial \theta_p}\big |_{p_I^*, p_F^*}<0$). The magnitude of this effect is shaped by the intensity of quality competition (i.e., $\theta_q$). Specifically, as explained below, a higher $\theta_q$ amplifies the effect of reduced data volume.

When $\theta_q$ increases, the provider is incentivized to raise the fine-tuning price, as shown in Lemma \ref{lemma: competition affect fine-tuning price}. This increase in $p_F^*$ amplifies the loss of fine-tuning revenue caused by the reduced data volume resulting from intensified price competition ($p_F^* \frac{\partial V_i(p_I,p_F)}{\partial\theta_p}\big |_{p_I^*, p_F^*}<0$). Consequently, the effect of reduced data volume becomes more pronounced as $\theta_q$ rises. When $\theta_q$ is relatively high (i.e., $\theta_q>\tilde{\theta}_q$), the negative effect of reduced data volume outweighs the positive effect of increased demand, ultimately leading to a decrease in the provider's profits.
This outcome, in which the provider's profits decrease in response to intensified downstream price competition within the context of the AI supply chain, contrasts with established insights from traditional supply chains involving manufacturers and retailers \citep{tyagi1999effects, YENIPAZARLI2021102428}. The key distinction lies in a novel mechanism we identify in the AI supply chain: Intensified price competition reduces data usage for fine-tuning, which in turn diminishes both fine-tuning and inference revenues for the provider. \looseness=-1



In contrast to the effects of price competition, Proposition~\ref{prop: provider profits with competition}\ref{item: quality competition increases provider profits} indicates that intensified quality competition always benefits the foundation model provider. To analyze this effect, we also apply the chain rule to analyze $\frac{d \pi_M}{d \theta_q}\big|_{p_I^*, p_F^*}$ and present the simplified expression below, with the complete derivation available in the Online Appendix:
\begin{equation*}
\frac{d \pi_M}{d \theta_q}\Bigg|_{p_I^*, p_F^*}=\sum_{i=1}^2 \biggl(
\frac{\partial {\pi}_M}{\partial V_i} 
\frac{\partial V_i(p_I,p_F)}{\partial \theta_q}
+\frac{\partial {\pi}_M}{\partial p_i}\frac{\partial p_i(p_I,p_F)}{\partial \theta_q}\biggr)\Bigg|_{p_I^*, p_F^*},
\end{equation*}
where $\frac{\partial {\pi}_M}{\partial V_i}\big |_{p_I^*, p_F^*}=p_I^*+p_F^*-2c_F V_i^*$ and $\frac{\partial {\pi}_M}{\partial p_i}\big|_{p_I^*, p_F^*}=-p_I^*$. As quality competition intensifies, firms increase their data usage for fine-tuning to enhance product quality ($\frac{\partial V_i(p_I,p_F)}{\partial \theta_q}\big |_{p_I^*, p_F^*}>0$). This rise in fine-tuning demand not only boosts fine-tuning revenue ($p_F^*\frac{\partial V_i(p_I,p_F)}{\partial \theta_q}\big |_{p_I^*, p_F^*}>0$) but also increases inference demand due to improvements in product quality, leading to higher inference revenue ($p_I^*\frac{\partial V_i(p_I,p_F)}{\partial \theta_q}\big |_{p_I^*, p_F^*}>0$). Although the increase in data volume raises the provider's compute costs, the net effect on profits remains positive because ($(p_I^*+p_F^*-2c_F V_i^*) \frac{\partial V_i(p_I,p_F)}{\partial \theta_q}\big |_{p_I^*, p_F^*}>0$). Additionally, as product quality improves, firms are incentivized to raise product prices, leading to increased double marginalization ($\frac{\partial p_i(p_I,p_F)}{\partial \theta_q}\big |_{p_I^*, p_F^*}>0$). This price increase can reduce the inference demand and thereby reduce the provider's profits ($-p_I^*\frac{\partial p_i(p_I,p_F)}{\partial \theta_q}\big |_{p_I^*, p_F^*}<0$). However, the positive impact on profitability from increased data volume always outweighs the negative impact from higher prices. Consequently, the provider's profits always increase as quality competition intensifies in the downstream market.

Next, we examine how increases in the intensity of price and quality competition, as measured by higher values of $\theta_p$ and $\theta_q$, respectively, affect downstream firms' profits and summarize our findings in the following corollary.

\begin{corollary}
[Impact of Downstream Competition on Firms’ Profits]
\label{coro: firm profits with competition}
\quad
\begin{enumerate}[label=\alph{enumi}., ref=\alph{enumi}]
	\item Intensified price competition may lead to an increase in downstream firms' profits.
	\item Intensified quality competition always results in a decrease in downstream firms' profits.
\end{enumerate}
\end{corollary}

Corollary~\ref{coro: firm profits with competition} reveals that intensified price competition in the downstream market can actually increase the profits of downstream firms. This finding challenges the conventional view that greater price competition always reduces profitability \citep{tyagi1999effects}. The intuition behind this result is as follows. On the one hand, consistent with conventional wisdom, intensified price competition prompts firms to lower their product prices, which typically reduces revenues. On the other hand, in response to increased price competition, the upstream provider can lower both fine-tuning and inference prices, as shown in Lemmas~\ref{lemma: competition affect fine-tuning price} and~\ref{lemma: competition affect inference price}. These price reductions reduce the costs of downstream firms to use provider services. When the savings from these cost reductions exceed the revenue losses from lower product prices, downstream firms can experience an increase in profits.
 
Corollary \ref{coro: firm profits with competition} further indicates that intensified quality competition always reduces downstream firms' profits, which aligns with conventional expectations. As quality competition intensifies, firms must invest more in product quality to remain competitive, resulting in higher fine-tuning costs. Additionally, as established in Lemmas \ref{lemma: competition affect fine-tuning price} and \ref{lemma: competition affect inference price}, the foundation model provider responds by increasing both fine-tuning and potentially inference prices, further diminishing downstream profitability. 

Proposition~\ref{prop: provider profits with competition} and Corollary~\ref{coro: firm profits with competition} offer key insights for foundation model providers, downstream firms, and policymakers. 
First, these results help both foundation model providers and downstream firms anticipate how their profits may evolve when pro-competitive policies are implemented in the downstream market. 
Such projections, given the resulting shifts in competitive dynamics, provide valuable guidance for strategic planning \citep{ForcastProfitIsImportant}.
Second, these findings assist policymakers by clarifying how different pro-competitive interventions affect industry participants, thereby supporting more informed decision-making in AI supply chain regulation. 
In particular, policymakers should be aware of the potential downsides of pro-quality-competitive policies.
While Proposition~\ref{prop: competition} demonstrates that such policies are robust in improving consumer surplus, Corollary~\ref{coro: firm profits with competition} reveals that they simultaneously reduce profitability for downstream firms, complicating their practical implementation. 
In contrast, Proposition~\ref{prop: provider profits with competition} and Corollary~\ref{coro: firm profits with competition} suggest that pro-price-competitive policies may benefit both foundation model providers and downstream firms. 
To assess whether such policies can deliver a ``win–win–win'' outcome that benefits consumers, downstream firms, and foundation model providers, we examine this possibility in the following corollary.

\begin{corollary}
[Win–Win–Win Outcome of Pro-price-competitive Policy]
\label{coro: win-win-win}
There exist cases in which a pro-price-competitive policy increases consumer surplus while simultaneously boosting the profits of both the foundation model provider and downstream firms.
\end{corollary}

Corollary~\ref{coro: win-win-win} indicates that a pro-price-competitive policy can generate a ``win–win–win'' outcome, benefiting all stakeholders. 
In contrast, a pro-quality-competitive policy typically increases consumer surplus at the cost of reduced firm profitability. 
This highlights the potential advantage of prioritizing price competition over quality competition, suggesting that policymakers should weigh the broader benefits of pro-price-competitive policies, even though pro-quality-competitive policies are always effective in enhancing consumer surplus.

In addition to pro-competitive policies, we now examine another key policy instrument—compute subsidies—to explore how subsidizing access to computing power affects the profitability of both the foundation model provider and downstream firms. The main result is summarized in Proposition~\ref{prop: compute subsidies and profits}.

\begin{proposition}
[Impact of Compute Subsidies on Profits]
\label{prop: compute subsidies and profits}
\indent 
\begin{enumerate}[label=\alph{enumi}., ref=\alph{enumi}]
\item Downstream firms' profits decrease due to compute subsidies only when $c_F>\hat{c}_F$.
\item The foundation model provider's profits always increase when compute subsidies are provided.
\end{enumerate}
For brevity, the expression for \( \hat{c}_F \) is provided in the Online Appendix.
\end{proposition}

Proposition~\ref{prop: compute subsidies and profits} highlights that while compute subsidies reduce compute costs for providers, they can unexpectedly decrease downstream firms' profits under certain conditions. This counter-intuitive outcome arises from two opposing sets of effects.
On the one hand, compute subsidies reduce the fine-tuning price, as discussed following Lemma~\ref{lemma: subsidy increase consumer surplus}. This price reduction lowers the fine-tuning fees paid by downstream firms, potentially boosting their profits.
Additionally, the lower price encourages firms to use more data for fine-tuning, thereby improving product quality. The resulting quality improvement drives the demand for the product, which in turn increases downstream firms' revenues.
On the other hand, the increase in data volume has two negative consequences for downstream firms.
First, it directly raises their data preprocessing costs.
Second, improvements in product quality stimulate inference demand, prompting the provider to raise the inference price. As a result, downstream firms face higher inference fees. Their profits decline when the increase in data preprocessing costs and inference fees combined exceeds the savings from reduced fine-tuning fees and the additional revenue from increased product sales. This outcome is especially likely when the coefficient of compute costs (i.e., $c_F$) is relatively high. A high $c_F$ results in a high fine-tuning price, which in turn limits data volume (i.e., $V_i$).
As a result, the savings in fine-tuning fees induced by the lower fine-tuning price (i.e., $V_i \cdot \Delta p_F$) due to compute subsidies are minimal, and the positive effect of the subsidy is outweighed by the higher data preprocessing costs and inference fees, ultimately reducing downstream firms’ profits.

In contrast to downstream firms' profits, the foundation model provider's profits always increase when compute subsidies are introduced. This is because subsidies directly reduce the compute costs for the fine-tuning service. In addition, they lower the fine-tuning price, encouraging downstream firms to invest more in product quality through fine-tuning. As product quality improves, the demand for inference services rises. Overall, the combination of reduced compute costs and increased inference revenue enhances the foundation model provider's profits. Since Proposition~\ref{prop: compute subsidies and profits} indicates that subsidies can enhance the profitability of both the provider and downstream firms, this regulatory instrument may lead to a "win-win-win" outcome. We summarize the condition for this outcome in the following corollary.

\begin{corollary}[Win–Win–Win Outcome of Compute Subsidies]
\label{coro: win-win-win of compute subsidies}
If $c_F < \hat{c}_F$, the provision of compute subsidies boosts consumer surplus and increases the profits of both the foundation model provider and downstream firms.
\end{corollary}

Corollary~\ref{coro: win-win-win of compute subsidies} indicates that compute subsidies can create a ``win–win–win'' outcome, benefiting all stakeholders when compute costs are relatively low. 
The mechanism behind this outcome has been discussed following Lemma~\ref{lemma: subsidy increase consumer surplus} and Proposition~\ref{prop: compute subsidies and profits}. 
Specifically, compute subsidies reduce the provider’s compute costs (which benefits the provider), prompting the provider to lower the fine-tuning price. 
This reduction, in turn, reduces the costs of downstream firms (which benefits downstream firms) and strengthens their incentives to co-create and improve quality, ultimately benefiting consumers. 

Our findings have important policy implications for the AI supply chain.
Corollary~\ref{coro: win-win-win of compute subsidies} and Proposition~\ref{prop: subsidy} suggest that policymakers should consider implementing compute subsidies when compute costs are relatively low (i.e., $c_F < \min \{\hat{c}_F, \hat{c} - 2c_V\}$). 
In such cases, compute subsidies not only effectively enhance consumer surplus but also simultaneously improve the profitability of both providers and downstream firms, making them a desirable policy intervention. 

Beyond this specific recommendation, synthesizing the findings of Sections~\ref{sec: Consumer Surplus and Policy Implications} and \ref{section: profits of foundation model provider and firms} reveals a fundamental trade-off in policy design: balancing the robustness of consumer benefits with the feasibility of achieving gains throughout the AI supply chain. This underscores the need for carefully calibrated interventions. We identify two policy classes, each with distinct strengths and limitations:
First, pro-quality-competitive policies are guaranteed to effectively enhance consumer surplus, but they cannot achieve a ``win-win-win'' outcome because they reduce downstream firms' profitability.
Second, pro-price-competitive policies and compute subsidies follow a different pattern. While they are not guaranteed to effectively enhance consumer surplus, they have the potential to create a "win-win-win" outcome that benefits all stakeholders.
This potential for broad buy-in across the AI supply chain makes these policies more practical to implement, even if their benefits to consumers are conditional.
From this perspective, policymakers face a trade-off: Policies that robustly benefit consumers may fail to benefit participants across the entire AI supply chain, whereas policies that can benefit both upstream and downstream participants may only deliver consumer benefits under certain conditions.

\section{Effect of Declining Compute Costs on Profits and Consumer Surplus}
\label{section: Impact of declining compute costs}
Beyond assessing the immediate impacts of regulatory policies, it is crucial to account for how their adoption may need to evolve over time in response to the rapidly changing landscape of AI markets. A key factor driving this shift is the ongoing reduction in compute costs, largely due to advancements in graphics processing units (GPUs), which are central to fine-tuning AI models \citep{GPUIsComputingPower,FineTuningNeedGPU}. To better understand these dynamics, we conduct a comparative statics analysis to examine how declining compute costs affect consumer surplus, as well as the profits of both the foundation model provider and downstream firms. We also examine how regulatory policies should adapt to maintain consumer-surplus-enhancing outcomes.

Intuitively, lower compute costs should enhance profitability across the AI ecosystem and benefit consumers \citep{ImpactOfDecliningGPUCosts}. 
However, prior research suggests that reductions in IT costs can reshape competitive dynamics, potentially eroding profits for firms in settings with only downstream participants \citep{demirhan2005information}.
In our setting, which involves an AI supply chain comprising foundation model providers, downstream firms, and consumers, the competitive dynamics are more complex. As a result, it is unclear how declining compute costs will impact profits at different layers of the supply chain, and whether the impacts on providers and downstream firms will differ.
Additionally, in the AI supply chain, where downstream firms co-create value with foundation model providers through fine-tuning, these effects from declining compute costs can influence firms' incentives to invest in quality, which in turn affects consumer surplus.
This underscores the need for a careful examination of how changes in compute costs impact both consumer surplus and the profitability of providers and downstream firms within the AI supply chain. The findings of this analysis are presented in Proposition~\ref{prop: profits with compute costs}.

\begin{proposition}
[Impact of Compute Costs]
\label{prop: profits with compute costs}
As the coefficient of compute costs (i.e., $c_F$) declines:
\begin{enumerate}[label=\alph{enumi}., ref=\alph{enumi}]
        \item Downstream firms' profits decrease if $c_F>\hat{c}_F$ and increase otherwise; \label{item: firm profits with compute costs}
	\item The foundation model provider's profits and consumer surplus always increase.\label{item: provider profits with compute costs}
\end{enumerate}

For brevity, the expression for \( \hat{c}_F \) is provided in the Online Appendix. 
\end{proposition}

Contrary to intuition, our results reveal that a decrease in $c_F$ can reduce firms' profits. This counter-intuitive effect arises from the following mechanism. Using the chain rule, we analyze $\frac{d  \pi_i}{d c_F}\big|_{p_I^*,p_F^*}$ as follows:
\begin{equation*}
    \frac{d  \pi_i}{d c_F}\Bigg|_{p_I^*,p_F^*}=
    \biggl(
    \frac{\partial {\pi}_i}{\partial p_F}\frac{d p_F^*}{d c_F}
    +\frac{\partial {\pi}_i}{\partial p_I}\frac{d p_I^*}{d c_F}
    +\frac{\partial {\pi}_i}{\partial V_i}\frac{d V_i^*}{d c_F}
    +\frac{\partial {\pi}_i}{\partial p_i}\frac{d p_i^*}{d c_F}
    \biggr)\Bigg|_{p_I^*, p_F^*},
\end{equation*}
where the expression for ${\pi}_i$ is given in Equation \eqref{SPNE pi}.

As $c_F$ decreases, two opposing forces shape its impact on downstream firms' profits. On the one hand, a lower $c_F$ reduces the marginal cost of fine-tuning for the foundational model provider, leading to a decrease in the fine-tuning price ($-\frac{d p_F^*}{d c_F}<0$). This reduction reduces the firms' fine-tuning expenses, thereby increasing their profits ($-\bigl(\frac{\partial {\pi}_i}{\partial p_F}\frac{d p_F^*}{d c_F} \bigr)\big|_{p_I^*, p_F^*}>0$). On the other hand, lower compute costs also affect profits through changes in the inference price, firms' product prices, and the data volume used for fine-tuning. The combined effect of these three channels is negative, i.e., $-\bigl(\frac{\partial {\pi}_i}{\partial p_I}\frac{d p_I^*}{d c_F}+\frac{\partial {\pi}_i}{\partial V_i}\frac{d V_i^*}{d c_F}+\frac{\partial {\pi}_i}{\partial p_i}\frac{d p_i^*}{d c_F}
\bigr)\big|_{p_I^*,p_F^*}<0$. 
This occurs because a lower fine-tuning price intensifies quality competition, which prompts firms to increase their investment in quality, which raises their expenditures (i.e., data preprocessing costs and fine-tuning fees) and directly reduces their profits. 
Furthermore, improvements in product quality increase consumer demand, which, in turn, raises the provider's inference demand. 
In response, the provider may increase the inference price charged to the firms, further eroding their profits. 
Together, these negative effects, combined with the positive effect from the lower fine-tuning price, generate a non-monotonic relationship between $c_F$ and downstream firms’ profitability. 
When $c_F$ is relatively high, the high compute cost of fine-tuning prompts the provider to set a correspondingly high fine-tuning price, which ultimately results in a low data volume. 
Given this low data volume, the reduction in the fine-tuning price resulting from declining compute costs leads to only a small profit gain for firms ($\frac{\partial \pi_i}{\partial p_F}\big|_{p_I^*, p_F^*} = -V_i^*$), resulting in a weak positive effect. 
Consequently, the negative effects dominate, causing the profits of downstream firms to decrease as compute costs decline.

These findings contribute to the literature on the implications of declining IT costs in competitive markets. Early work by \citet{demirhan2005information} shows that declining IT costs generally benefit firms by enabling greater investment, which in turn enhances product quality and increases profits. However, \citet{demirhan2007strategic} find that when the switching costs between competing firms' products are sufficiently high, declining IT costs can reduce firm profitability. This occurs because lower IT costs prompt firms to engage in aggressive quality competition, thereby eroding their profits. This result is consistent with our finding that firms may also be worse off under declining compute costs. However, our model introduces a different mechanism behind this outcome. In \citet{demirhan2007strategic}, overinvestment arises from intertemporal competition, in which high switching costs lock consumers into future purchases, incentivizing firms to overinvest in the initial period to secure long-term demand. 
In contrast, in our model, the pricing behavior of the foundation model provider plays a central role. Specifically, as compute costs fall, the provider lowers the fine-tuning price, encouraging firms to increase their investment in product quality. This mechanism reflects the co-creation dynamics of the AI supply chain, which, to our knowledge, has not been identified in the existing literature.

Proposition~\ref{prop: profits with compute costs}\ref{item: provider profits with compute costs} further shows that the foundation model provider's profits always increase as $c_F$ decreases. This occurs because lower compute costs ($c_F$) reduce the provider's expenses for delivering fine-tuning services, which in turn reduces the fine-tuning price. The lower price incentivizes downstream firms to increase data volumes and enhance product quality, thereby boosting inference demand and increasing inference revenue. Taken together, the reduction in fine-tuning costs and the increase in inference revenue result in higher provider profits as compute costs decline.
Similarly, consumer surplus increases as compute costs decline. 
This is because lower compute costs lead to a reduction in the provider’s fine-tuning price, which enhances downstream firms’ incentives to engage in the co-creation process and improve product quality. 
The resulting improvements in product quality ultimately increase consumer surplus.

While Proposition~\ref{prop: profits with compute costs} demonstrates that consumer surplus increases as compute costs decline, it remains unclear whether regulatory interventions that are effective under current cost structures will continue to be appropriate as costs decrease further. 
To address this, we examine how declining compute costs affect the conditions for effective policy, as outlined in Proposition~\ref{prop: competition} and Proposition~\ref{prop: subsidy}. 
The results are summarized in the following corollary.

\begin{corollary}[Impact of Declining Compute Costs on Policies]
\label{coro: declining compute costs on policies}
As compute costs decline, (i) compute subsidies are more likely to be effective in enhancing consumer surplus, (ii) the pro-price-competitive policy becomes less likely to be effective in enhancing consumer surplus, and (iii) the pro-quality-competitive policy remains consistently effective.
\end{corollary}

Recall that compute subsidies are effective only when compute and data preprocessing costs are relatively low, while the pro-price-competitive policy is effective only when these costs are relatively high. 
Thus, Corollary~\ref{coro: declining compute costs on policies} suggests that as compute costs decline, compute subsidies may shift from being ineffective to effective. 
This implies that policymakers should consider adopting this policy in the future, even if it appears ineffective under current cost conditions. 
In contrast, as compute costs continue to fall, the pro-price-competitive policy may shift from being an effective tool for enhancing consumer surplus to one that harms it. 
Therefore, policymakers should exercise caution with this policy and consider phasing it out once compute costs fall to a relatively low level. 
Finally, the declining trend in compute costs does not significantly affect the effectiveness of the pro-quality-competitive policy. This policy consistently increases consumer surplus, regardless of compute cost levels, making it a robust tool for enhancing consumer surplus.
Overall, since compute costs are expected to continue declining, our findings highlight the importance of ongoing monitoring and adaptive policymaking.

\section{Conclusion}
Rapid innovation in AI has brought regulatory concerns to the forefront, driven by fears that companies might exploit AI technologies to consolidate market power and capture an outsized share of consumer surplus. This study examines these concerns from a systemic perspective, focusing on the emerging AI supply chain and the interactions between upstream and downstream participants.
We analyze two types of policies within this supply chain and evaluate their effectiveness. To protect consumers, pro-competitive policies are commonly used; however, the specific approach to promoting competition is critical. Promoting downstream quality competition consistently enhances consumer surplus, whereas promoting downstream price competition is beneficial only when either compute costs or data preprocessing costs (or both) are high.
Importantly, these cost conditions relate to both the foundation model provider's compute costs and the downstream firms' data preprocessing costs, highlighting the co-creation nature of the AI supply chain.
Furthermore, we introduce a new class of policy aimed at enhancing surplus within the AI supply chain: compute subsidies. Our analysis shows that compute subsidies, which can effectively enhance consumer surplus when both the compute and data preprocessing costs are low, complement pro-price-competitive policies. We also examine how these policies affect the participants in the AI supply chain. We find that while pro-quality-competitive policies tend to reduce downstream firms' profits, they increase the profits of the foundation model provider. In contrast, both pro-price-competitive policies and compute subsidies can boost consumer surplus and increase profits for both providers and downstream firms, creating a ``win-win-win'' outcome where all stakeholders benefit. Finally, we explore the implications of declining compute costs. As these costs continue to fall, pro-price-competitive policies may shift from being an effective tool for enhancing consumer surplus to a harmful one that diminishes it. In contrast, compute subsidies may transition from an ineffective policy to an effective one as compute costs decline.

\subsection{Theoretical Contributions}
Our study makes significant theoretical contributions. Specifically, we take a crucial first step in characterizing regulation within the AI supply chain by systematically analyzing the interactions between upstream and downstream participants. This is particularly important given the limited theoretical research on regulation in traditional supply chain contexts.
Furthermore, our study highlights the distinctive features of the AI supply chain, setting it apart from existing research on traditional supply chains and software markets (which is relevant because fine-tuning and inference of foundation models are delivered as software-enabled services).
First, unlike traditional supply chains, the development of fine-tuned models involves co-creation between foundation model providers and downstream firms. Additionally, the provider generates revenue through two streams: inference and fine-tuning, whereas manufacturers in traditional supply chains typically rely on wholesale revenue. 
Second, the co-creation process in the AI supply chain differs from other well-known co-creation processes among firms. In traditional settings, co-creation typically occurs independently across participants, with each party contributing autonomously to enhance the final product. Examples include vendor–client collaborations for enterprise systems (e.g., ERP), cases in the computer industry where one firm develops the central processor while another develops the operating system, and scenarios in the smartphone industry where one firm designs the device and operating platform, while others develop applications for it \citep{demirezen2020two,puyang2025vertical, gupta2023worse, ceccagnoli2012cocreation}.
In contrast, fine-tuning foundation models involves a mutually dependent form of co-creation. The foundation model provider not only supplies the base model but also provides the computational infrastructure needed for downstream firms to use their domain-specific data to improve the model. As downstream firms contribute more data to enhance model performance, they increase the compute demand, thus raising the provider's compute costs. This interdependence in the co-creation process is largely absent from traditional co-creation settings.
Third, the customization process in the AI supply chain differs significantly from that in traditional software markets. In conventional settings, software providers sell licenses to users, who then customize and deploy the software on their own infrastructure, usually for a fixed one-time fee \citep{ma2015analyzing,zhang2020cloud}. In contrast, fine-tuned AI models are typically hosted on the provider’s cloud infrastructure and offered through a pay-as-you-go pricing model. 
Fourth, this deployment mechanism and pricing structure place both fine-tuning and inference services within the broader SaaS category. However, these services differ significantly from traditional SaaS in terms of customization flexibility. While traditional SaaS applications often offer limited customization options \citep{SaaSLimitedCustomization, zhang2020cloud}, fine-tuning services enable users to adapt AI models with much greater flexibility to meet specialized needs.

In the AI supply chain, we also identify several novel mechanisms through which policy interventions affect consumers and other participants, arising from its unique co-creation structure.
First, pro–price-competitive policies in downstream markets increase the provider’s incentive to engage in co-creation, leading the provider to lower the fine-tuning price. This, in turn, improves product quality and enhances consumer surplus. In contrast, pro–quality-competitive policies reduce the provider’s co-creation incentives, leading to an increase in the fine-tuning price, negatively impacting product quality.
Second, while pro–price-competitive policies reduce product prices and are therefore expected—according to conventional wisdom—to increase provider profits through reduced double marginalization, they also diminish downstream firms’ incentives to improve quality. This results in less domain-specific data being used for fine-tuning, reducing fine-tuning and inference revenue, and ultimately lowering the provider’s profits.
Third, the introduction of compute subsidies or a decline in compute costs strengthens the provider's incentives to engage in co-creation. This, in turn, encourages the provider to lower the fine-tuning price, motivating downstream firms to invest more in product quality. This mechanism reflects the unique co-creation dynamics of the AI supply chain, which are not captured in existing studies on the implications of declining compute costs \citep{demirhan2005information,demirhan2007strategic}. \looseness=-1


\subsection{Policy Implications}
From a practical standpoint, our research provides valuable insights for policymakers seeking to enhance consumer surplus within the AI supply chain. 
First, we demonstrate that pro-price-competitive policies and compute subsidies are not universally effective; their effectiveness depends on the underlying cost conditions (as shown in Proposition~\ref{prop: competition} and Proposition~\ref{prop: subsidy}). This allows policymakers to tailor interventions to specific market conditions. Importantly, these two policies can complement each other,  as each tends to be effective in situations where the other is not. For instance, in downstream markets dealing with complex, sensitive, and high-stakes data—such as healthcare and finance \citep{WorstDataIndustry}—where firms incur high data preprocessing costs, pro-price-competitive policies may be ineffective or even harmful to consumer welfare, whereas compute subsidies are likely to provide substantial benefits. In contrast, in markets with relatively low data preprocessing costs, such as e-commerce, where retailers typically use structured data formats for product details, offers, pricing, and reviews, making the data easier to preprocess \citep{StructuredDataEasyProcess, StructuredDataECommerce2}, pro-price-competitive policies can effectively enhance consumer surplus, whereas compute subsidies may have a more limited impact.

Second, many governments already provide compute subsidies at varying rates. For example, Beijing offers subsidies covering up to 30\% of compute costs, while Shenzhen covers up to 50\%. Our results show that excessively high subsidy rates can be counterproductive: Their effectiveness diminishes and may even become negative once the subsidy rate exceeds a certain threshold (as shown in Corollary~\ref{coro: subsidy}). Therefore, our results offer clear guidance to policymakers on effective subsidy design, suggesting that a moderate subsidy rate leads to the greatest net increase in consumer surplus. 

Third, we find that pro-price-competitive policies and compute subsidies are generally more viable and more likely to gain support across the AI supply chain than pro-quality-competitive policies (as shown in Proposition~\ref{prop: provider profits with competition}). This is because both pro-price-competitive policies and compute subsidies have the potential to create a ``win–win–win'' outcome for consumers, downstream firms, and foundation model providers. While pro-quality-competitive policies always improve consumer welfare, they consistently reduce downstream firms’ profits and thus cannot deliver such a mutually beneficial outcome—an important consideration for regulatory adoption.

Fourth, our analysis highlights the importance of dynamic and adaptive policymaking. As GPUs improve and compute costs decline, the effectiveness of specific policy tools may evolve (as shown in Corollary~\ref{coro: declining compute costs on policies}): pro-price-competitive policies may shift from effective to ineffective, while compute subsidies may become effective after initially being ineffective. Policymakers should therefore closely monitor cost trends and adjust their regulatory strategies accordingly, using our framework to navigate evolving technological conditions. 

Finally, as compute costs continue to fall, our findings suggest that policymakers should increasingly emphasize compute subsidies (as shown in Corollary~\ref{coro: win-win-win of compute subsidies} and Proposition~\ref{prop: subsidy}).
In future low-cost environments, compute subsidies not only effectively increase consumer surplus but also help create a robust ``win–win–win'' outcome, making them a particularly promising tool for fostering socially beneficial AI supply chains.


%
%
%



\bibliographystyle{informs2014} 
\bibliography{main.bib} 

\end{document}